\pdfoutput=1
\pdfoutput=1
\documentclass[acmtog]{acmart}
\acmSubmissionID{407}
\AtBeginDocument{%
  \providecommand\BibTeX{{%
    \normalfont B\kern-0.5em{\scshape i\kern-0.25em b}\kern-0.8em\TeX}}}

\definecolor{MyRed}{rgb}{0.65,0.07,0.09}

\definecolor{MyGreen}{rgb}{0.18,0.55,0.09}

\usepackage{wrapfig}
\usepackage{diagbox}
\usepackage{verbatim}
\usepackage{amssymb}
\usepackage{latexsym}
\usepackage{lineno}
\usepackage{xspace}
\usepackage{color}
\graphicspath{{figures/}}
\usepackage{amsmath,bm}
\usepackage{psfrag}
\usepackage{mathtools}

\theoremstyle{definition}

\usepackage{multirow, tabularx}
\newcolumntype{Y}{>{\centering\arraybackslash}X}
\usepackage{capt-of}%
\usepackage{array, boldline, makecell, booktabs}

\newcommand{\Mod}[1]{\ (\mathrm{mod}\ #1)}
\usepackage{array}
\usepackage{pifont}
\usepackage{psfrag}
\usepackage{makecell}
\usepackage{algorithm}
\usepackage{threeparttable,booktabs}
\usepackage{graphicx}
\usepackage{subcaption}
\usepackage{flushend}
\usepackage{stfloats} 
\usepackage{appendix}
\usepackage[noend]{algpseudocode}

\setcopyright{acmlicensed}
\acmJournal{TOG}
\acmYear{2024} \acmVolume{43} \acmNumber{6} \acmArticle{} \acmMonth{12}\acmDOI{10.1145/3687916}

\usepackage{xcolor}
\usepackage{graphicx}

\newcommand{\bo}[1]{\textcolor{blue}{[Bo: #1]}}

\newif\ifverbose
\verbosetrue

\ifverbose

\newcommand{\vb}[1]{\textcolor{red}{#1}}
\else

\newcommand{\vb}[1]{}
\fi

\usepackage{url}
\usepackage{xcolor}
\definecolor{newcolor}{rgb}{.8,.349,.1}

\newcommand{\revise}[1]{\textcolor{black}{#1}}

\begin{document}

\author{Zhiqi Li}
\email{zli3167@gatech.edu}
\affiliation{
\institution{Georgia Institute of Technology}
\country{USA}
}

\author{Duowen Chen}
\email{dchen322@gatech.edu}
\affiliation{
\institution{Georgia Institute of Technology}
\country{USA}
}

\author{Candong Lin}
\email{lincandong@outlook.com}
\affiliation{
\institution{Georgia Institute of Technology}
\country{USA}
}

\author{Jinyuan Liu}
\email{jinyuan.liu.gr@dartmouth.edu}
\affiliation{
\institution{Dartmouth College}
\country{USA}
}

\author{Bo Zhu}
\email{bo.zhu@gatech.edu}
\affiliation{
\institution{Georgia Institute of Technology}
\country{USA}
}


\title{Particle-Laden Fluid on Flow Maps}

\begin{abstract}
We propose a novel framework for simulating ink as a particle-laden flow using particle flow maps. Our method addresses the limitations of existing flow-map techniques, which struggle with dissipative forces like viscosity and drag, thereby extending the application scope from solving the Euler equations to solving the Navier-Stokes equations with accurate viscosity and laden-particle treatment. Our key contribution lies in a coupling mechanism for two particle systems, coupling physical \revise{sediment} particles and virtual flow-map particles on a background grid by solving a Poisson system. We implemented a novel path integral formula to incorporate viscosity and drag forces into the particle flow map process. Our approach enables state-of-the-art simulation of various particle-laden flow phenomena, exemplified by the bulging and breakup of suspension drop tails, torus formation, torus disintegration, and the coalescence of sedimenting drops. In particular, our method delivered high-fidelity ink diffusion simulations by accurately capturing vortex bulbs, viscous tails, fractal branching, and hierarchical structures.
\end{abstract}



\begin{teaserfigure}
\centering%
\includegraphics[width=1.\textwidth]{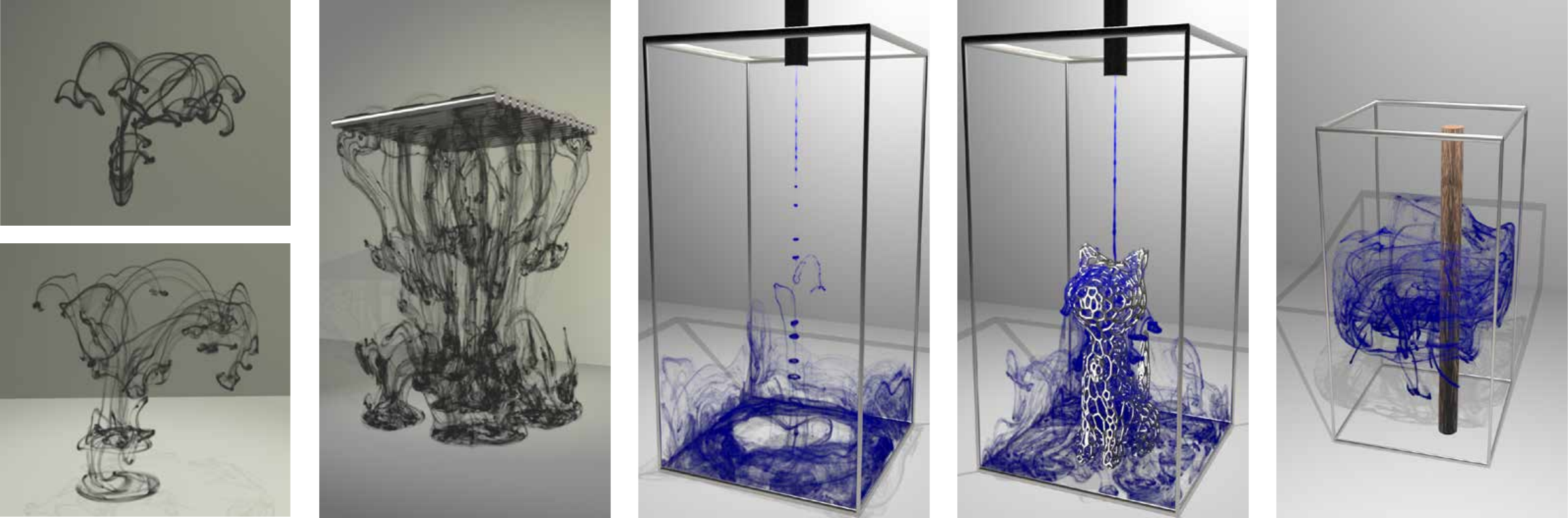}
\caption{Examples of ink diffusion phenomena simulated by our method. Five simulation snapshots are shown from left to right: the interaction of two ink drops forming ink rings and filaments; the formation of complex patterns as ink flows down a porous plate; the bulging, breakup, and torus formation of the tail of a dripped ink drop; the interaction between ink and a hollow cat; and the ink diffusion stirred by a chopstick.}
 \label{fig:teaser}
\end{teaserfigure}

\maketitle

\section{Introduction}
The diffusion process of ink in water exhibits fascinating flow characteristics. Starting from a localized, high-concentration region, the ink diffuses into its fluid environment by forming incipient blobs, filament tails, fractal branches, and chandelier-like structures, exhibiting a tremendous mix of pattern similarities and flow complexities that have attracted experimental observation, mathematical study, and numerical simulation from scientists across different disciplines. Despite the visually complicated flow process during ink diffusion, its underlying governing physics is surprisingly simple: As studied in previous literature on fluid mechanics (e.g., see \cite{nitsche1997break}), ink diffusion in water can be described as a typical particle-laden flow process, where ink particles are dispersed within the fluid medium and interact with the fluid flow through the drag force produced on each particle's granular shape. The concentrated blobs, elongated tails, and fractal branches form due to the flow instability among vorticity, viscosity, and particle-laden drags.

However, despite the elegance of the physics picture, devising a first-principle method to simulate ink diffusion remains a challenging problem in computational physics and computer graphics. \revise{Pioneering efforts to simulate ink diffusion studied the vorticity-viscosity interaction and produced simulation results of one layer of branching with vortex particles in the early 1990s \cite{nitsche1997break}}, which, to the best of our knowledge, is the first work simulating ink diffusion from the perspective of particle-laden flow. This work motivated future works in the area (e.g.,  \revise{\cite{bosse2005numerical,machu2001coalescence,walther2001three}}). In computer graphics, researchers have repurposed the standard grid-based smoke simulator with vorticity confinement \cite{fedkiw2001visual} to a particle-laden flow setting for ink diffusion, which reproduced impressive \revise{ink branching phenomena} \revise{\cite{sagong2015simulating}}. In the recent work of \cite{padilla2019bubble}, the authors simulated ink chandeliers as vortex filaments with varying thicknesses and achieved notable multi-layer chandelier structures. Despite these inspiring successes, a versatile simulation framework that can naturally handle the vortical development and structural evolution by directly solving particle-laden flow in an arbitrary incompressible flow environment remains to be developed. In particular, the challenge of directly simulating vorticity-viscosity interaction to spontaneously emerge the ink blobs, filaments, branches, and structures by directly solving the Navier-Stokes (NS) equations with \revise{sediment} particles remains an essential problem to be addressed.

To this end, we propose a novel and versatile framework to simulate ink as a particle-laden flow. Our method is based on the recent line of works on solving incompressible flow with flow maps (e.g., \cite{nabizadeh2022covector,deng2023fluid,li2024lagrangian,zhou2024eulerian,li2023garm}). 
Though these flow-map schemes have demonstrated their extraordinary capabilities in preserving accurate vortical structures --- which would be particularly beneficial for ink diffusion --- they inherently suffer from limitations in tackling dissipative forces such as viscosity and drag. \textit{To date, no flow-map paradigms can accurately solve for viscosity, let alone its interaction with vorticity}. Mathematically, this is due to the lack of a mathematical foundation for calculating viscosity (or any forces other than pressure) along each particle's trajectory. 

To complete this missing piece, which will further unleash the flow-map method's notable vorticity-preserving ability in a broader scope of scenarios, in particular, scenarios where vorticity-viscosity interaction plays an important role, we propose a novel particle flow-map simulation method to solve the \textbf{full Navier-Stokes equations} with laden-particle interactions. The key technical novelties we delivered include two parts: 
On the one hand, we have devised two particle systems, one for physical, \revise{sediment} particles (with mass and momentum) to track mass transport, and the other for the virtual, flow map particles (massless) to evolve vortical structures. \revise{Sediment} particles and flow-map particles can exchange information via a background grid during the Poisson solve for incompressibility. 
On the other hand, we devised a novel path integral formula and a flexible flow map adaption strategy to incorporate viscosity and drag forces into the previously purely geometric mapping process. This Euler-to-NS enhancement, in conjunction with its particle-laden coupling, will enable the state-of-the-art simulation of ink diffusion by accurately capturing the flow details during the entire development of ink diffusion, including the vortex bulb, viscous tails, fractal branching, and hierarchical structures. 

The main contributions of our approach include:
\begin{itemize} 
    \item A novel path integral form for the Navier-Stokes equations with accurate viscosity and drag treatments;
    \item A novel mechanism to couple long-range flow maps, short-range forces, and a short-range projection with variable-coefficient Poisson equations;
    \item A unified Eulerian-Lagrangian particle-laden flow solver that can facilitate the state-of-the-art ink diffusion simulations exhibiting complex vorticity-viscosity interaction.
\end{itemize}

\section{Related Work}
\paragraph{Eulerian-Lagrangian Methods.}
Hybrid methods in fluid simulation merge the strengths of both Lagrangian and Eulerian approaches, resulting in more versatile and robust systems. Since the pivotal work on PIC \cite{harlow1962particle} and FLIP \cite{brackbill1986flip} was introduced to the graphics community by \citet{zhu2005animating}, hybrid Eulerian-Lagrangian representations have become prevalent in fluid simulations \cite{deng2022moving,raveendran2011hybrid,zhu2010creating}. The Material Point Method (MPM), which extends the concepts of PIC/FLIP, has been employed to simulate a range of continuum behaviors, such as collision and fracture \cite{stomakhin2013material}, viscoplasticity \cite{yue2015continuum}, magnetization \cite{sun2021material}, solid-fluid interaction \cite{fang2020iq}, and sedimentation \cite{gao2018animating}. Ongoing research has improved the accuracy of transfers between particles and grids, addressing issues such as unconserved vorticity \cite{jiang2015affine,fu2017polynomial}, displacement discontinuity \cite{hu2018moving}, and volume conservation \cite{qu2022power}.
\paragraph{Particle-Laden Flow.}
Particle-laden flow refers to the movement of a fluid that contains suspended particles, such as dust in the air \cite{xiu2020numerical}, sediment in water \cite{gao2018animating}, or droplets in a gas \cite{xu2023breakup}. The accurate representation of such flows involves modeling the interactions between particles and fluids, including drag \cite{sagong2015simulating}, dispersion \cite{xu2011interactive}, and phase transition \cite{vignesh2022cluster}. While the particles are typically treated as Lagrangian when coupling with the fluid  \cite{xu1997numerical}, they can behave like a secondary fluid and be modeled through a two-phase flow solver \cite{kartushinsky2016eulerian} or a multiscale continuous approach \cite{idelsohn2022multiscale}. Inks made of fine-scale pigment particles can interact with fluids and exhibit complex diffusion patterns, such as bulging and breakup \cite{machu2001coalescence}, branching \cite{thomson1886v}, and other topological transitions. Numerous experiments have been conducted to explore these processes through high-speed photography \cite{thoroddsen2008high} and particle image velocimetry \cite{zhou2022experimental}, capturing particle speed and size \cite{wijshoff2018drop}, viscosity \cite{krainer2019effect}, and temperature \cite{lee2004ink}. Ink simulations include vorticity confinement \cite{sagong2015simulating}, vortex particles \cite{walther2001three}, and vortex filaments \cite{padilla2019bubble}.

\paragraph{Flow Map Methods.}
Flow map, a geometric representation of spatial time slice, was widely used to reduce diffusion errors presented in interpolations and advections. Such a method was first introduced in \cite{wiggert1976numerical} in computational physics and later brought to the graphics community in \cite{hachisuka2005combined}. The idea of mapping between simulation time slices connected by tracing virtual particles between frames was explored in \cite{sato2017long, sato2018spatially, qu2019efficient}, but the mapping happened between velocity fields. From the introduction of \revise{covector fluids} \cite{nabizadeh2022covector} to the graphics community, the benefit of flow map aiding the advection of covectors proved its superiority in vorticity concentrated \revise{phenomena}. Flow map methods for covectors were improved by \cite{deng2023fluid} based on a neural buffer to facilitate a bidirectional map, which has been further improved with particles \cite{sancho2024impulse, zhou2024eulerian,li2024lagrangian} to reduce memory consumption. 


\section{Physical Model}
\paragraph{Naming Convention}
We distinguish fluid and sediment using superscripts $f$ and $s$. We denote sediment particles with subscript $p$. We use $\textbf{u}$ and $\textbf{v}$ to denote fluid and sediment velocities.

\subsection{Fluid}
Based on \cite{gao2018animating,nielsen2013two,sun2016sedifoam}, the fluid momentum and mass conservation equations are
\begin{equation}\label{equ:original_mass_equation}
\begin{cases}
    \frac{\partial(\epsilon^f\rho^f\mathbf{u})}{\partial t}+\nabla\cdot(\epsilon^f\rho^f\mathbf{u}\otimes \mathbf{u})=-\epsilon^f\nabla p^f+\epsilon^f\rho^f \mathbf{g}+\mu\epsilon^f\Delta \textbf{u}+\mathbf{f}^{f}_{\text{drag}},\\
    \frac{\partial(\epsilon^f\rho^f)}{\partial t}+\nabla\cdot(\epsilon^f\rho^f\mathbf{u})=0,
\end{cases}
\end{equation}
where $\epsilon^f$, $\rho^f$, $\revise{\mu}$, $\mathbf{u}$, $p^f$, $\mathbf{g}$ and $\mathbf{f}_{\text{drag}}^f$ are the fluid volume fraction, fluid intrinsic density, \revise{fluid viscosity coefficient}, fluid velocity, fluid pressure, the gravitational constant and the fluid drag force density respectively. Assuming constant fluid density $\rho^f$ and combined with the mass conservation equation, the fluid momentum equation can be reformulated as
\begin{equation}\label{equ:fluid_momentum}
        \frac{\partial \mathbf{u}}{\partial t}+(\mathbf{u}\cdot\nabla)\mathbf{u}=-\frac{1}{\rho^f}\nabla \revise{p^f}+\textbf{g}+\frac{\mu}{\rho^f}\Delta\textbf{u}+\frac{1}{\rho^f\epsilon^f}\textbf{f}_{\text{drag}}^f.
\end{equation}
\revise{Assuming constant sediment density $\rho^s$ and} combined with mass equation for sediment $\frac{\partial(\epsilon^s\rho^s)}{\partial t}+\nabla\cdot(\epsilon^s\rho^s \textbf{v})=0 $ and with relationship of volume fraction $\epsilon^f+\epsilon^s=1$, the fluid incompressibility condition is obtained as
\begin{equation}\label{equ:mixture_incompressibility_constraint}
    \begin{aligned}
        \nabla\cdot (\epsilon^f\textbf{u}+\epsilon^s\textbf{v})=0,
    \end{aligned}
\end{equation}
where $\epsilon^s$, $\rho^s$, and $\textbf{v}$ are the sediment volume fraction, sediment intrinsic density, and sediment velocity respectively. 

\subsection{Sediment}
Similar to \cite{sun2016sedifoam}, the sediment is described by a collection of sediment particles following Newton's Law.  For each particle $p$, it holds
\begin{equation}\label{equ:sediment_dynamics}
    \frac{d\mathbf{v}_p}{dt}=\revise{\left(\frac{\rho^s-\rho^f}{\rho^s}\right)}\mathbf{g}+\frac{1}{m_p}\mathbf{f}_{\text{drag},p},
\end{equation}
where $\mathbf{v}_p$ and $m_p=\rho^sV_p$ denote velocity and mass of \revise{sediment} particles respectively.  Here, $V_p=\frac{4}{3}\pi r_p^3$ and $r_p$ is the radius of a sediment particle.  For viscous laden flow like ink, we use Stokes' Law to calculate the drag force $\mathbf{f}_{\text{drag},p}$, given by:
\begin{equation}\label{equ:drag_force_sediment}
    \mathbf{f}_{\text{drag},p}=6\pi \mu r_p(\textbf{u}(\textbf{x}_p,t)-\revise{\textbf{v}_p}),
\end{equation}
where $\textbf{x}_p$ denote the position of particle $p$. By Newton's third law, for a fluid microelement occupying the space $\mathcal{V}$, the fluid drag force density can be calculated as
\begin{equation}\label{equ:drag_force_fluid}
    \textbf{f}_{\text{drag}}^f=-\frac{1}{|\mathcal{V}|}\sum_{\textbf{x}_p\in \mathcal{V}} \textbf{f}_{\text{drag},p},
\end{equation}
where $|\mathcal{V}|$ denote the volume of space $\mathcal{V}$.

\newcolumntype{z}{X}
\newcolumntype{s}{>{\hsize=.25\hsize}X}
\begin{table}[t]
\centering\small
\begin{tabularx}{0.47\textwidth}{scz}
\hlineB{3}
Notation & Type & Definition\\
\hlineB{2.5}
\hspace{12pt}$*^f$ & all & Fluid property\\
\hlineB{1}
\hspace{12pt}$*^s$ & all & \revise{Sediment} particle property\\
\hlineB{1}
\hspace{12pt}$*_p$ & all & property on sediment particle $p$\\
\hlineB{1}
\hspace{12pt}$*_q$ & all & property on fluid particle $q$\\
\hlineB{1}
\hspace{12pt}$*_i$ & all & fluid or sediment property on the grid $i$\\
\hlineB{1}
\hspace{12pt}$r$ & scalar & Current time\\
\hlineB{1}
\hspace{12pt}$s$ & scalar & Initial time\\
\hlineB{1}
\hspace{12pt}$s'$ & scalar & One time step before the current time\\
\hlineB{1}
\hspace{12pt}$\Phi_{a\to b}$ & vector & Forward flow map from time $a$ to $b$\\
\hlineB{1}
\hspace{12pt}$\Psi_{b\to a}$ & vector & Backward flow map from time $b$ to $a$\\
\hlineB{1}
\hspace{12pt}$\mathcal{F}_{a\to b}$ & matrix & Jacobian of forward flow map $\Phi_{a\to b}$\\
\hlineB{1}
\hspace{12pt}$\mathcal{T}_{b\to a}$ & matrix & Jacobian of backward flow map $\Psi_{b\to a}$\\
\hlineB{1}
\hspace{12pt}$\Gamma_{a\to b}$ & vector & Covector-based integration of fluid force from time $a$ to  $b$\\
\hlineB{1}
\hspace{12pt}$\mathbf{u}_{a\to b}^M$ & vector & \textbf{Mapped} velocity at time $b$ by covector flow map from time $a$\\
\hlineB{1}
\hspace{12pt}$\mathbf{u}_{a\to b}^A$ & vector & \textbf{Advected} velocity at time $b$ by particle velocity from time $a$\\
\hlineB{3}
\end{tabularx}
\caption{Summary of important notations used in the paper.}
\label{tab: notation_table}
\end{table}

\section{Particle Flow Map for Laden Flow}
\subsection{Flow Map Theory}

\setlength{\intextsep}{0pt}%
\setlength{\columnsep}{10pt}%

\begin{wrapfigure}[16]{r}{0.2\textwidth}
\centering
\vspace{-0.2cm}
\includegraphics[width=0.2\textwidth]{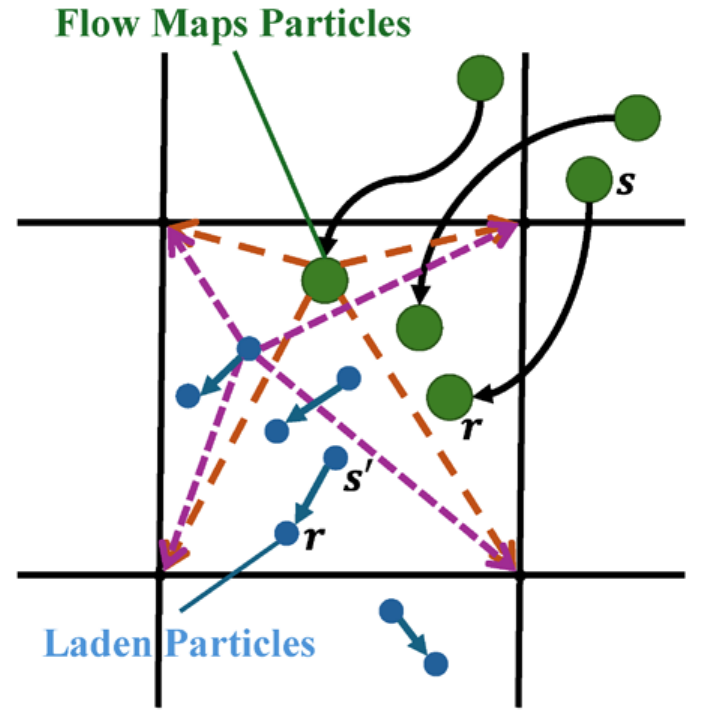}
\vspace{-0.75cm}\caption{Flow Maps are tracked by fluid particles $q$ and long-range mapped velocity is calculated, while \revise{sediment} particles $p$ advect short-range velocity step-by-step.  They interact via the background grid with interpolation and Poisson solve.}\label{fig:fm}
\end{wrapfigure}

\paragraph{Flow Map}  During the motion of fluid with velocity $\mathbf{u}_t(\mathbf{x}) = \mathbf{u}(\mathbf{x},t)$, $\forall t > 0$, consider an arbitrary fluid particle moves from the position $\mathbf{x}_s \in \Omega_s$ at the initial time $s$ to the position $\mathbf{x}_r \in \Omega_r$ at the current time $r$, where $\Omega_s$ and $\Omega_r$ represent the domain at times $s$ and $r$ respectively.  The positions $\mathbf{x}_s$ and $\mathbf{x}_r$ can be associated using the forward flow map $\Phi_{s \to r}:\Omega_s\to \Omega_r$ and the backward flow map $\Psi_{r \to s}:\Omega_r\to \Omega_s$, satisfying $\Phi_{s \to r}(\mathbf{x}_s) = \mathbf{x}_r$ and $\Psi_{r\to s}(\mathbf{x}_r) = \mathbf{x}_s$. The Jacobian matrices of the flow maps $\Phi_{s \to r}$ and $\Psi_{r \to s}$ are denoted as $\mathcal{F}_{s \to r}(\mathbf{x}_s) = \frac{\partial \Phi_{s \to r}(\mathbf{x}_s)}{\partial \mathbf{x}_s}$ and $\mathcal{T}_{r \to s} = \frac{\partial \Psi_{r \to s}(\mathbf{x}_r)}{\partial \mathbf{x}_r}$ respectively. According to \cite{deng2023fluid}, for any time $t$, $\mathcal{F}_{s \to t}$ and $\mathcal{T}_{t \to s}$ satisfy the evolution equations:
\begin{equation}\label{equ:advection_FT}
    \begin{aligned}
        \frac{D\mathcal{F}_{s\to t}}{Dt}&=\nabla \mathbf{u}(\mathbf{x}) \mathcal{F}_{s\to t},\\
\frac{D\mathcal{T}_{t\to s}}{Dt}&= -\mathcal{T}_{t\to s}\nabla \mathbf{u}(\mathbf{x}),\\
    \end{aligned}
\end{equation}
where $\mathcal{F}_{s \to t}$ and $\mathcal{T}_{t \to s}$ are abbreviations for $\mathcal{F}_{s \to t}(\Psi_{t \to s}(\mathbf{x}))$ and $\mathcal{T}_{t \to s}(\mathbf{x})$, respectively.  \revise{The} $\Psi_{r\to s}$, as the mapping $\Psi_{r\to s}: \Omega_r \to \Omega_s$, induces a pullback of scalar fields  $(\Psi_{r\to s})^*: \Omega_s^*\to \Omega_r^* $ and a pullback of covector fields  $(\Psi_{r\to s})^*: \mathfrak{X}^*(\Omega_s) \to \mathfrak{X}^*(\Omega_r) $ where $\Omega_r^*$ is the space of scalar field on $\Omega_r$ and $\mathfrak{X}^*(\Omega_r)$ is the space of covector field on $\Omega_r$ (the pullback of $\Phi_{s\to t}$ is similar, see \cite{Crane2013Exteriorcalculus}). 

\paragraph{Particle Flow Map}  To solve incompressible and inviscid fluid, \citet{li2024lagrangian} and \citet{zhou2024eulerian} use particles as flow maps to calculate the path integral form (see \cite{li2024lagrangian} for details):
\begin{equation}\label{equ:path_intergral_last}
    \begin{aligned}
        \mathbf{u}_q(r)=\mathcal{T}_{r\to s,q}^T\mathbf{u}_q(s)-\nabla \Lambda_{r\to s,q},
    \end{aligned}
\end{equation}
where the subscript $q$ denotes quantities carried by the particle $q$, $\Lambda_{r\to s,q}=\int_s^r\lambda\left(\Phi_{s\to \tau}(\Psi_{r\to s}(x_q(r))),\tau\right)d\tau$, $\lambda=p-\frac{1}{2}|u|^2$ and $x_q(r)$ is the position of $q$ at time $r$.  Based on \autoref{equ:path_intergral_last}, the map-projection process is used to solve the incompressible Euler equation:
\begin{enumerate}
    \item (\textbf{Long-Range Mapping}) Calculate the long-range mapped velocity on particles as: \({\mathbf{u}_{s\to r,q}^{M}} = {\mathcal{T}_{r\to s,q}}^T \mathbf{u}_{s,q}\);
    \item (\textbf{Long-Range Projection}) \revise{Solve the Poisson equation regarding $\Lambda_s^r$:} ${\nabla\cdot\nabla \Lambda_{s}^r =\nabla\cdot {\mathbf{u}_{s\to r}^{M}}}$ on the grid to obtain $\Lambda_{s}^r$.  Then carry out a projection as \(\mathbf{u}_{r} = {\mathbf{u}_{s\to r}^{M}} - \nabla \Lambda_{s}^r\) to ensure \(\mathbf{u}_{r}\) satisfies \(\nabla \cdot \mathbf{u}_r = 0\).    
\end{enumerate}
In this scheme, particles are used to track flow maps and compute mapped velocity, while the grid or other discrete structures can be used to perform projection. 

\begin{figure}[t]
 \centering
\includegraphics[width=1.0\linewidth]{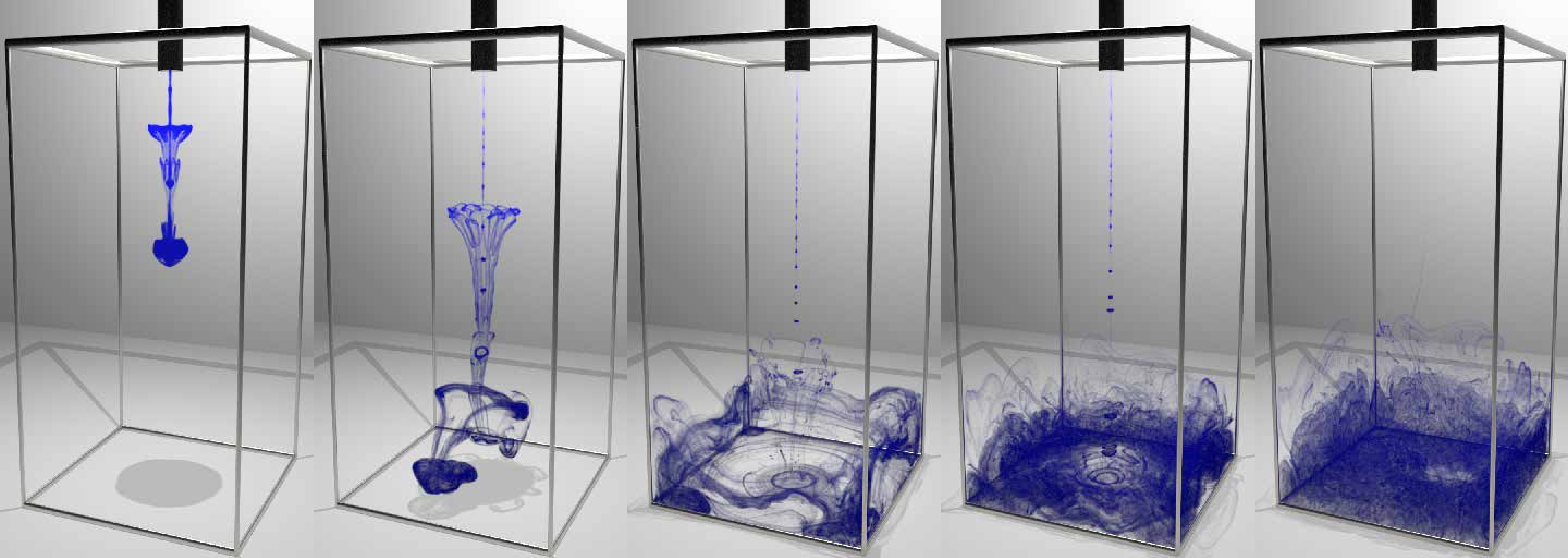}
\caption{Dripping. The long tail of a descending ink will bulge and break up into many suspension drops, which forms tori afterwards}
\label{fig:inject1}
\end{figure}

\subsection{Adapting Particle Flow Map to Laden Flow}


When applying the particle flow map method to laden flow, we encountered the following challenges:

\begin{enumerate}
    \item The integral form \autoref{equ:path_intergral_last} is derived from the Euler equation without considering dissipative forces, which makes the particle flow map method unable to account for forces other than fluid pressure.
    \item
    The particle flow map method cannot directly interact with sediment particles because it operates on a long-range mapped velocity $\mathbf{u}^M_{s\to r}$. In contrast, the velocity of sediment particles is described by step-by-step updates of short-range advected velocity $\mathbf{v}_p$. Although $\mathbf{u}_{r}$ can interact with $\mathbf{v}_p$, the interactions between the short-range fluid velocity $\mathbf{u}_{r}$ and the sediment particles' velocity $\mathbf{v}_p$ are short-range. They cannot be reflected in the long range.
    \item In the particle flow map method, the pressure integral $\Lambda_{s\to r}$ is solved by directly projecting the mapped velocity $\mathbf{u}^M_{s\to r}$. This approach is not applicable in sediment flow because the velocity in sediment flow needs to satisfy the incompressible constraint \autoref{equ:mixture_incompressibility_constraint}, which includes information about the sediment and the fluid's volume fraction. Leveraging this short-range information to enforce incompressibility with a Poisson equation for the long-range pressure $\Lambda_{s\to r}$ is difficult.
\end{enumerate}

\begin{figure}[t]
 \centering
 \includegraphics[width=.47\textwidth]{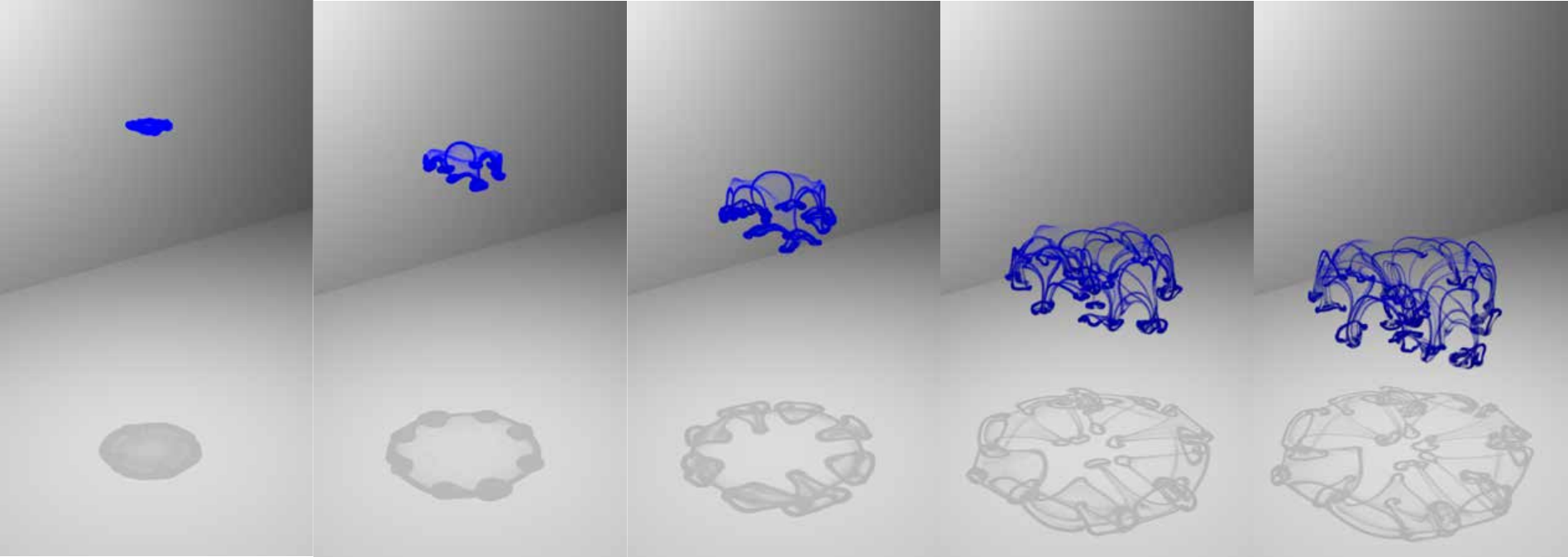}
 \caption{Ink Torus Breakup: Under drag force and viscous force, a ink torus disintegrate into several blobs, which deform into tori and disintegrate again,  resulting in a cascade process of blob deformations and breakups.}
 \label{fig:single_drop}
\end{figure}

\begin{figure}[t]
 \centering
 \includegraphics[width=.47\textwidth]{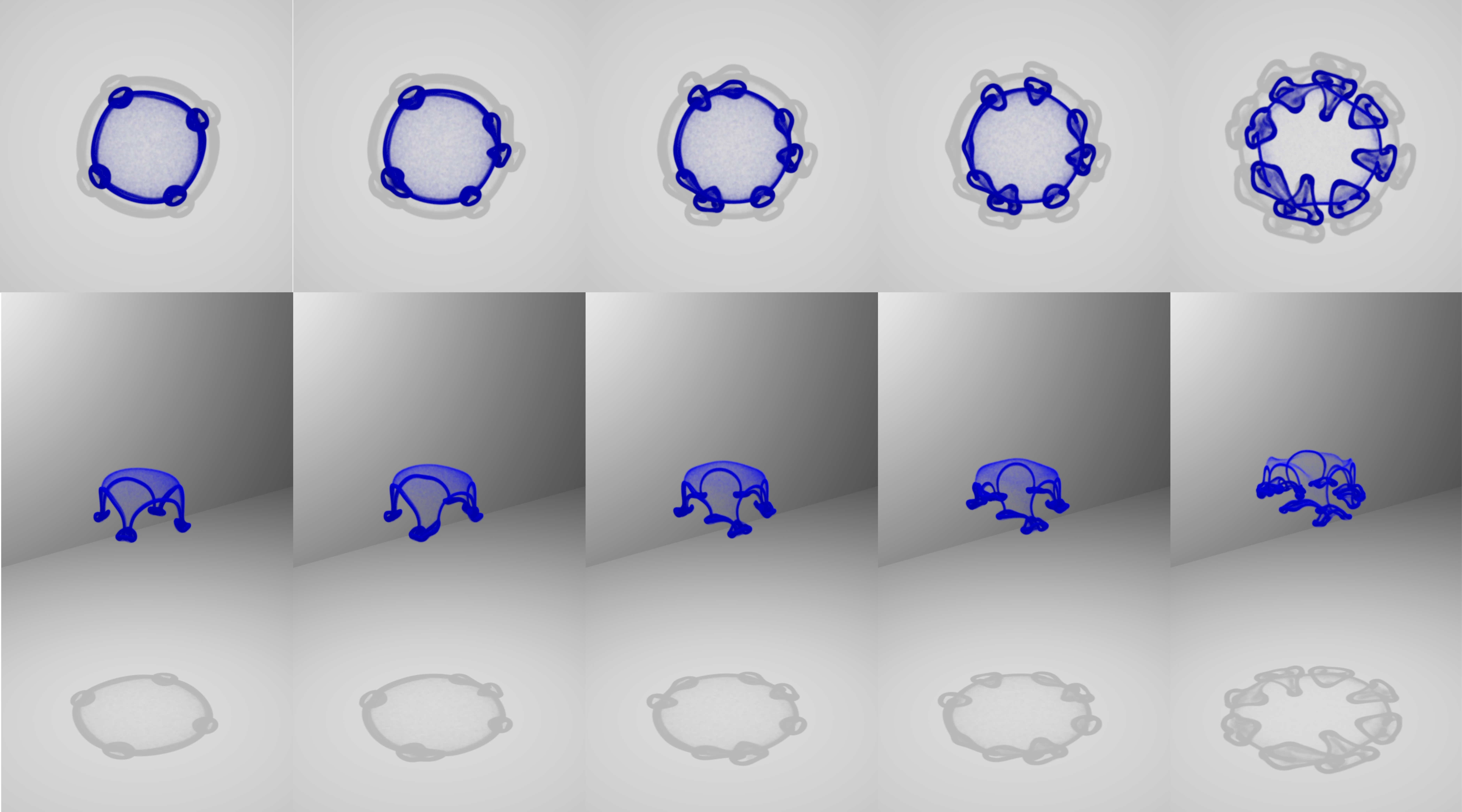}
 \caption{Ink Torus Breakup Under $Re= 16.5, 18, 19.5, 21.0, 30.0$: number of blobs disintegrated from a ink torus increase with Reynolds Number, under interaction of viscous force and vorticity. }
 \label{fig:multi_re}
\end{figure}

\subsection{Integration Form and Path Integral}
To address the first challenge, we propose a new integral formulation based on \autoref{equ:fluid_momentum} capable of handling forces other than pressure. For simplicity, we denote any force other than pressure as $\mathbf{\gamma}$, and rewrite \autoref{equ:fluid_momentum} as 
\begin{equation}\label{equ:fluid_momentum_gamma}
\frac{\partial \mathbf{u}}{\partial t}+(\mathbf{u}\cdot \nabla)\mathbf{u}=\mathbf{\gamma} -\frac{1}{\rho^f}\nabla p.
\end{equation}
By utilizing covector forms, we derived the integral form (see \autoref{sec:Covector_Derivation} for detailed derivation):
\begin{equation}
\begin{aligned}\label{equ:interation_our}
    \mathbf{u}(\mathbf{x},r)=\underbrace{{\mathcal{T}_{r\to s}^T(\mathbf{x}) \mathbf{u}_s(\Psi_{r\to s}(\mathbf{x}),s)}}_{\text{Mapping}}+ \underbrace{\mathcal{T}_{r\to s}^T(\mathbf{x})\Gamma_{s\to r}(\mathbf{x})}_{\text{Force Path Integral}},
\end{aligned}
\end{equation}
where $\Gamma_{s\to r}(\mathbf{x})$ is defined as
\begin{equation}\label{equ:definition_gamma}
    \Gamma_{s\to r}(\mathbf{x})=\revise{\int_s^r \mathcal{F}_{s\to \tau}^T\Psi_{r \to s}(\mathbf{x})\left(\gamma
-\frac{1}{\rho^f}\nabla \lambda\right)(\Phi_{s\to \tau}(\Psi_{r \to s}(\mathbf{x})),\tau)d\tau},
\end{equation}
where $\lambda=p-\frac{1}{2}\rho^f|\mathbf{u}|^2$ is the Lagrangian pressure with density $\rho^f$.

\begin{figure*}
\includegraphics[width=1.0\linewidth]{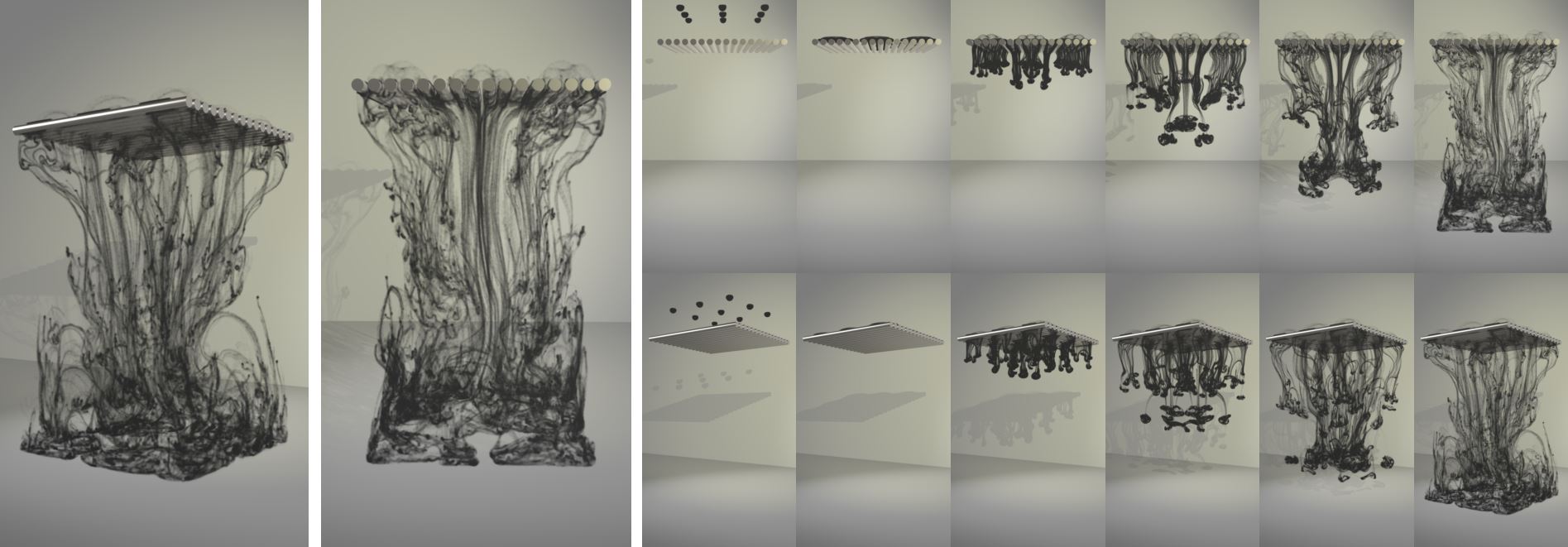}
\caption{Nine Ink
Drops Passing Porous Obstacle. Nine ink drops drip down from the gaps between cylinders, turning into many small falling drops.}
\label{fig:porus}
\end{figure*}

\begin{figure*}
\includegraphics[width=1.0\linewidth]{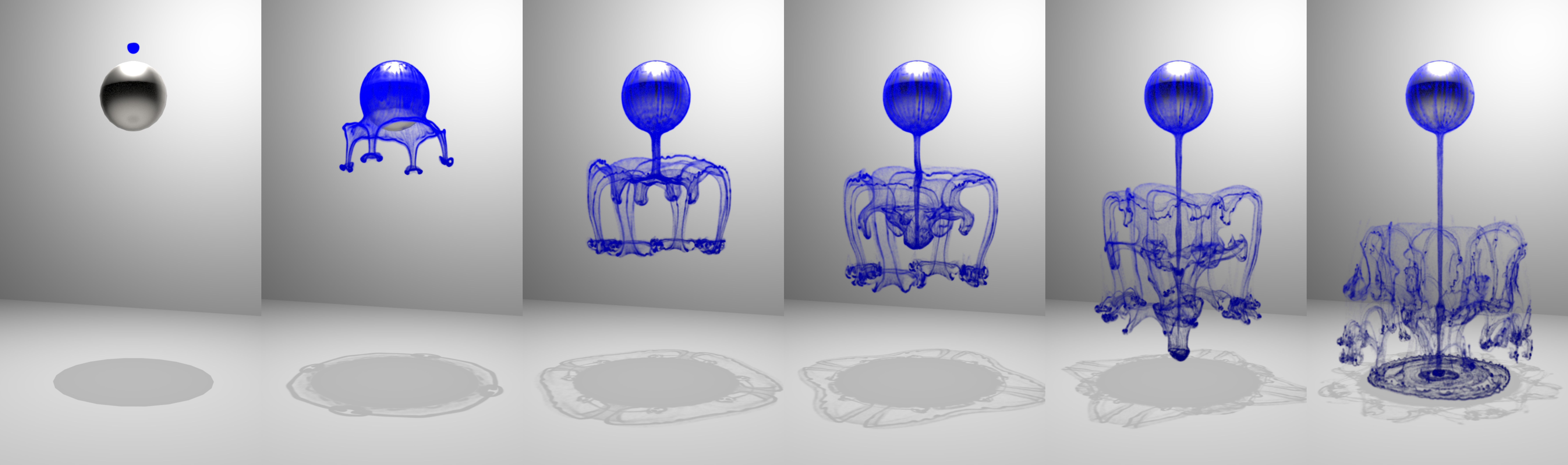}
\caption{One Ink
Drop Passing Sphere Obstacle. One ink drop drops on a sphere, some parts form small drops flowing along the surface while some parts form a gathered drop at sphere's bottom.}
\label{fig:small_sphere}
\end{figure*}

Note that the integral $\Gamma_{s\to r}$ can be reformulated into a path integral form along a Lagrangian trajectory. For a fluid particle $q$ with position $x_q(t)$ at time $t$, the force path integral term in \autoref{equ:interation_our} is the path integral of $\revise{\mathcal{F}_{s \to t}^T\left(\gamma-\frac{1}{\rho^f}\nabla \lambda\right)}$ along its trajectory from time $s$ to $r$, which is denoted as
\begin{equation}
\begin{aligned}
    \Gamma_{s\to r,q}=\int_s^r \mathcal{F}_{s\to \tau,q}^T\revise{\left(\gamma
-\frac{1}{\rho^f}\nabla \lambda\right)}(\mathbf{x}_q(\tau),\tau)d\tau,
\end{aligned}    
\end{equation}
where \( \mathcal{F}_{s\to \tau,q}^T=\mathcal{F}_{s\to \tau}^T(\mathbf{x}_q(s)) \) represents the Jacobian of the forward flow map from \( s \) to \( \tau \) along the particle trajectory, which could be carried on particles.  Then \autoref{equ:interation_our} can be reformulated as the path integral form as
\begin{equation}\label{equ:basic_path_integral_our}
\boxed{\mathbf{u}_{r,q} = \underbrace{{\mathcal{T}_{r\to s,q}}^T \mathbf{u}_{s,q}}_{\text{Mapping}} + \underbrace{\mathcal{T}^T_{r\to s,q }\Gamma_{s\to r,q}}_{\text{Force Path Integral}}.}
\end{equation}

\subsection{Mapped Velocity Conversion}\label{sec:conversion}
To address the second and third challenges, we adopt a similar idea of Long-Range Mapping Classical Projection (LMCP) proposed in \cite{li2024lagrangian}. We design a method based on \autoref{equ:basic_path_integral_our} to convert long-range mapped velocity $\mathbf{u}_{s\to r}^M$ into advected velocity $\mathbf{u}_{s'\to r}^A$, where $\mathbf{u}_{s'\to r}^A$ is the classical advected velocity advected by $\frac{D \mathbf{u}}{Dt}=0$ one time step from $s'$.  $s'$ is one time step before $r$. $\mathbf{u}_{s'\to r}^A$ is used to interact with sediment particle and interaction is \revise{accumulated} to long range by updating the path integrator, \revise{which allows us to address the second challenge}. Simultaneously, by performing projection on $\mathbf{u}_{s'\to r}^A$, we resolve the third challenge.

To present the formula for converting \( \mathbf{u}_{s \to r, q}^M \) to \( \mathbf{u}_{s' \to r, q}^A \), first, decompose \(\mathcal{T}_{r\to s}^T\Gamma_{s\to r,q}\) as 
\begin{equation}
    \begin{aligned}
        \mathcal{T}_{r\to s}^T\Gamma_{s\to r,q}=&\mathcal{T}_{r\to s}^T\Gamma_{s\to s',q} \\
         &+ \mathcal{T}_{r\to s}^T \int_{s'}^{r} \mathcal{F}^T_{s\to \tau,q}\revise{\left(\mathbf{\gamma}-\frac{1}{\rho^f}\nabla \lambda\right)}(\mathbf{x}_q(\tau),\tau)d\tau.
    \end{aligned}
\end{equation}
Here, the second part $\mathcal{T}_{r\to s}^T \int_{s'}^{r} \mathcal{F}^T_{s\to \tau,q}\revise{\left(\mathbf{\gamma}-\frac{1}{\rho^f}\nabla \lambda\right)}(\mathbf{x}_q(\tau),\tau)d\tau$ on the right-hand side of the above equation is equal to $\revise{\Big(}\gamma - \frac{1}{\rho^f}\nabla p + \frac{1}{2}\nabla |\mathbf{u}_{s'}|^2\revise{\Big)}(\mathbf{x}_q(r))\Delta t$ when using the Euler scheme to compute with the identity $\mathcal{T}_{r\to s}^T\mathcal{F}_{s\to r}^T=I$. Utilizing the relationship \(\mathbf{u}_{r, q} = \mathbf{u}_{s \to r, q}^M + \mathcal{T}_{r \to s, q}^T \Gamma_{s \to r, q} = \mathbf{u}_{s'\to r,q}^A+\revise{\left(\mathbf{\gamma}\Delta t-\frac{1}{\rho^f}\nabla p\right)}(\mathbf{x}_q(r))\), we can derive the formula for converting \(\mathbf{u}_{s' \to r, q}^M\) to \(\mathbf{u}_{s' \to r, q}^A\) as 

\begin{equation}\label{equ:long_to_advect}
\boxed{\mathbf{u}_{s'\to r,q}^A = \mathbf{u}_{s\to r,q}^M +\mathcal{T}_{r\to s}^T\Gamma_{s\to s',q}+\frac{1}{2}(\nabla|\mathbf{u}_{s'}|^2)(\mathbf{x}_q(r))\Delta t.} 
\end{equation}

In the following discussion, we use $\mathbf{u}_{s'\to r}^A$ to represent the field indicated by $\mathbf{u}_{s'\to r,q}^A$ on all particles.
 


\begin{figure*}
\includegraphics[width=0.96\linewidth]{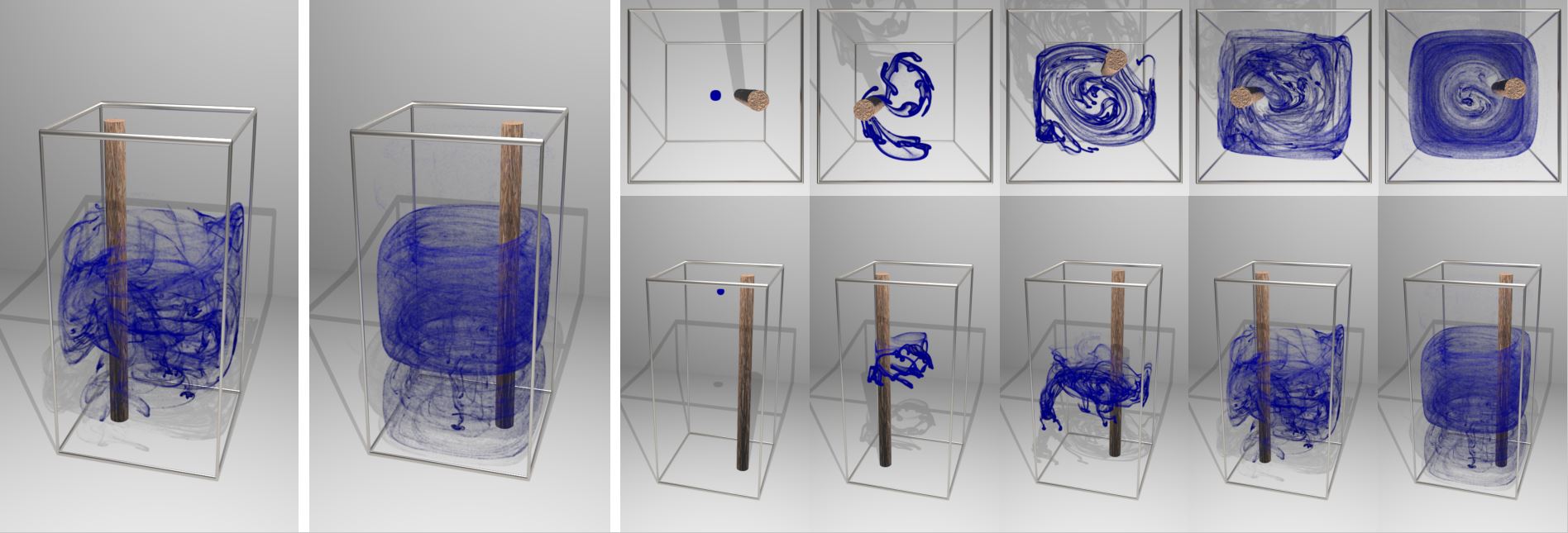}
\caption{ Chopstick
Stirring Ink. The stirred ink forms many layers and pulls out many fine threads, gradually spreading throughout the entire domain.}
\label{fig:stir}
\end{figure*}

\subsection{Force Effect}\label{sec:effect_of_force}

After converting $\mathbf{u}_{s\to r,q}^M$ to $\mathbf{u}_{s'\to r}^A$, we can incorporate forces $\mathbf{\gamma}$ and pressure $p$ in the fluid into $\mathbf{u}_{s'\to r}^A$ in a conventional manner. Once $\mathbf{\gamma}$ is computed, it is applied to $\mathbf{u}_{s'\to r}^A$ yielding $\mathbf{u}^* = \mathbf{u}_{s'\to r}^A + \mathbf{\gamma}\Delta t$. As $\mathbf{u}^*$ is derived from the classical advected velocity $\mathbf{u}_{s'\to r}^A$, the pressure $p$ can be directly obtained from the variable coefficient Poisson equation \autoref{equ:mixture_poisson} derived from \autoref{equ:mixture_incompressibility_constraint}. The projected velocity $\mathbf{u}_r$ is then computed as $\mathbf{u}_r = \mathbf{u}^* - \frac{1}{\rho_f}\nabla p$. Through this process, we address the third challenge.

Subsequently, it is necessary to accumulate the force $\mathbf{\gamma}$ and pressure $-\frac{1}{\rho_f}\nabla p$ into the path integrator $\Gamma_{s\to r,q}$.  With $\Delta \Gamma = \revise{\left(\gamma-\frac{1}{\rho^f}\nabla p\right)}$, $\Gamma_{s\to r,q}$ is updated by its definition \autoref{equ:definition_gamma} as

 
 \begin{equation}\label{equ:update_Gamma}
     \Gamma_{s\to r,q}=\Gamma_{s\to s',q}+\mathcal{F}_{s\to r_q}^T\left(\Delta \Gamma_q+\frac{1}{2}(\nabla |\mathbf{u}_{s'}|^2)(x_q(r))\right) \Delta t.
 \end{equation}
The updated $\Gamma_{s\to r,q}$ is used in the conversion between the mapped velocity and the advected velocity in the next step.

To be specific, we will substitute $\gamma$ with specific force and delve into detailed discussions regarding the drag force $\frac{1}{\rho^f\epsilon^f} \mathbf{f}_{\text{drag}}^f$, viscous force $\frac{\mu}{\rho^f}\Delta \mathbf{u}_{s'\to r}^A$, and gravity $\mathbf{g}$ appearing in \autoref{equ:fluid_momentum}. \revise{In the complete computation of laden flow, the effects of these three forces will be considered simultaneously according to the following discussion and summed up.}

\paragraph{Drag Force} First, let's consider when $\mathbf{\gamma}$ is the drag force. The drag force originates from the interaction between sediment particles and fluid, and since both the advected velocity $\mathbf{u}^A_{s'\to s}$ and $\mathbf{v}_q$ are short-range, they can interact with each other. Utilizing the formula in \autoref{equ:drag_force_sediment}, we compute the drag force $\mathbf{f}_{\text{drag},p}$ acting on sediment particles. Then, employing $\mathbf{f}_{\text{drag},p}$ and \autoref{equ:drag_force_fluid}, we calculate $\mathbf{f}^f_{\text{drag}}$, and accumulate it onto $\mathbf{u}^A_{s'\to s}$ to obtain $\mathbf{u}^*$. After projection of $\mathbf{u}^*$, $\mathbf{f}^f_{\text{drag}}$ needs to be accumulated onto long-range using $\Delta \Gamma = \revise{\left(\frac{1}{\rho^f\epsilon^f}\mathbf{f}^f_{\text{drag}}-\frac{1}{\rho^f}\nabla p\right)}$ and \autoref{equ:update_Gamma}. Through this process, we address the second challenge. The numerical computation process for $\mathbf{f}_{\text{drag},p}$ and $\mathbf{f}^f_{\text{drag}}$ is detailed in \autoref{sec:sediment_calculation}.

\paragraph{Viscous Force and Gravity}  When $\mathbf{\gamma}$ is the viscous force, we only need to compute $\Delta \mathbf{u}_{s'\to r}$. Then, we accumulate $\frac{\mu}{\rho^f}\Delta \mathbf{u}_{s'\to r}$ onto $\mathbf{u}^A_{s'\to r}$ to obtain $\mathbf{u}^*$. After projection, we can accumulate $\frac{\mu}{\rho^f}\Delta \mathbf{u}_{s'\to r}$ onto $\Gamma_{s\to r,q}$ using \autoref{equ:update_Gamma}. Gravity, on the other hand, can be directly accumulated onto $\mathbf{u}^A_{s'\to r}$ to obtain $\mathbf{u}^*$ and subsequently accumulated into $\Gamma_{s\to r,q}$ after projection.

\subsection{Conclusion of Algorithm}
\label{sec:time_split}
Based on \autoref{equ:long_to_advect} and \autoref{sec:effect_of_force}, we obtain our time-split scheme as:

\begin{enumerate}
    \item (\textbf{Long-Range Mapping}) Calculate the long-range mapped velocity as: \({\mathbf{u}_{s\to r,q}^{M}} = {\mathcal{T}_{r\to s,q}}^T \mathbf{u}_{s,q}\);
    \item (\textbf{\( \mathbf{u}_{s \to r, q}^M \) to \( \mathbf{u}_{s' \to r, q}^A\)}) Calculate $\frac{1}{2}(\nabla|\mathbf{u}_{s'}|^2)(\mathbf{x}_q(r))$ and convert $\mathbf{u}_{s \to r, q}^M$ to $\mathbf{u}_{s' \to r, q}^A$ based on \autoref{equ:long_to_advect};
     \item (\textbf{Exert Force}) Update sediment particle dynamics and calculate force like viscous force $\frac{\mu}{\rho^f}\Delta \mathbf{u}_{s'\to r}^A$ and drag force $\frac{1}{\rho^f\epsilon^f}\mathbf{f}_{\text{drag}}^f$ on the grid, as discussion in \autoref{sec:effect_of_force}.  Then calculate $\mathbf{u}^*$ as
     \begin{equation}\label{equ:compute_u_star}
         \mathbf{u}^*=\mathbf{u}_{s'\to r}^A+\revise{\left(\mathbf{g}+\frac{\mu}{\rho^f}\Delta \mathbf{u}_{s'\to r}^A+\frac{1}{\rho^f\epsilon^f}\mathbf{f}_{\text{drag}}^f\right)}\Delta t;
     \end{equation}
     \item (\textbf{Classical Projection}) Solve a variable coefficient Poisson equation derived from \autoref{equ:mixture_incompressibility_constraint} \cite{gao2018animating}:
     \begin{equation}\label{equ:mixture_poisson}
     \revise{
         -\frac{\Delta t}{\rho^f} \nabla \cdot(\epsilon^f \nabla p)=-\nabla \cdot(\epsilon^s \mathbf{v})-\nabla \cdot(\epsilon^f \mathbf{u}^*);}
     \end{equation}
     \item (\textbf{Update}) Project $\mathbf{u}^*$ by $\mathbf{u}_{r}=\mathbf{u}^*-\frac{1}{\rho^f}\nabla p$ and update $\Gamma_{s\to r,q}$ as \autoref{equ:update_Gamma}.
\end{enumerate}
In the above scheme, detailed calculations regarding \(\mathcal{T}_{s\to r,q}\), \(\mathcal{F}_{s\to r,q}\), \(\epsilon^f\), $\frac{1}{2}(\nabla|\mathbf{u}_{s'}|^2)(\mathbf{x}_q(r))$, etc. are discussed in \autoref{sec:numerial_calculation}.

\begin{figure*}
\centering
\begin{minipage}[t]{0.48\textwidth}
\centering
\includegraphics[width=1\linewidth]{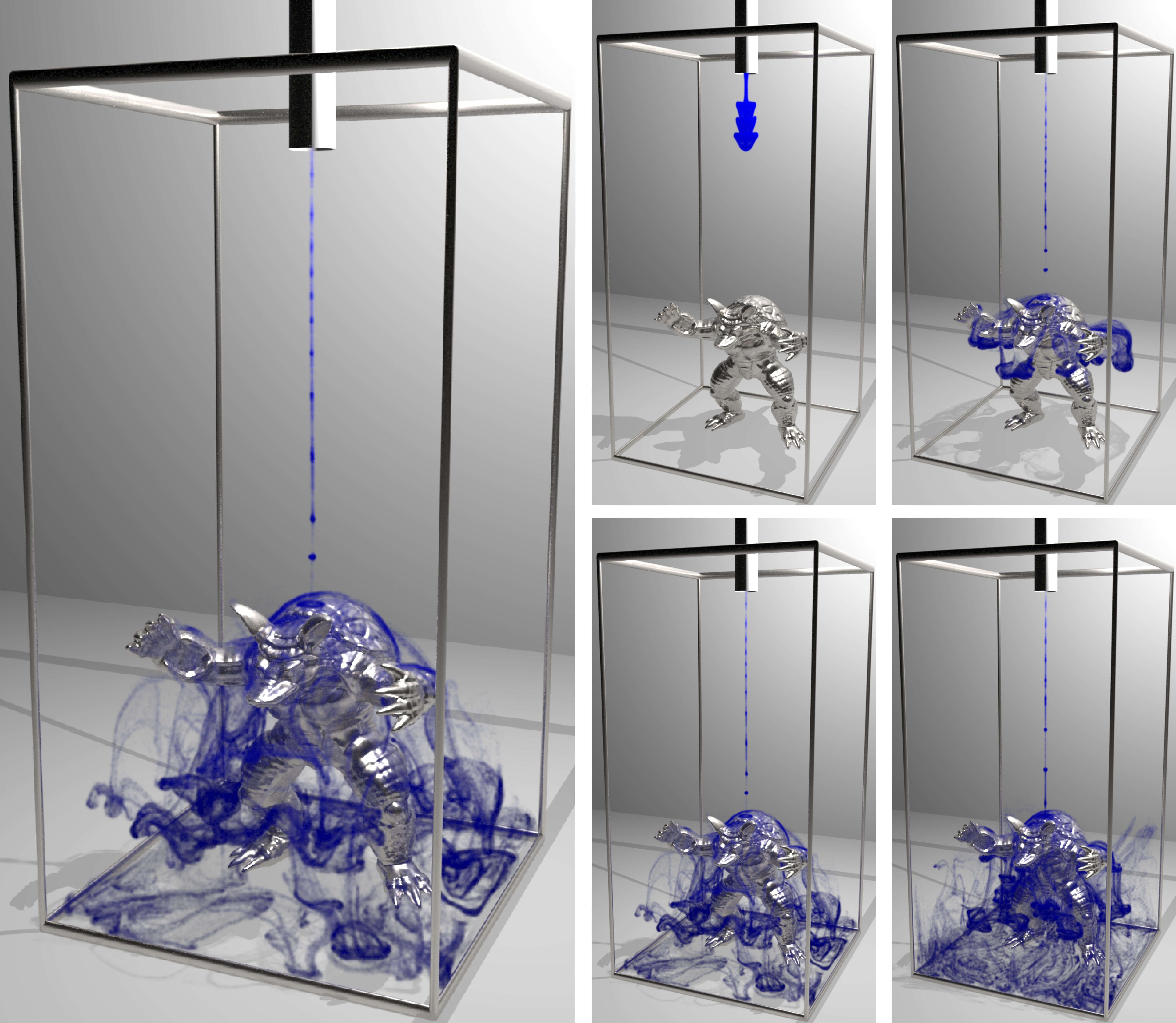}
\caption{Ink dropping onto an armadillo: once collides the ink splinters into numerous drops and coats the armadillo; descending drop tails bulge and break up into many small suspension drops.}
\label{fig:armadillo}
\end{minipage}
\hspace{0.2cm}
\begin{minipage}[t]{0.48\textwidth}
\centering
\includegraphics[width=1\linewidth]{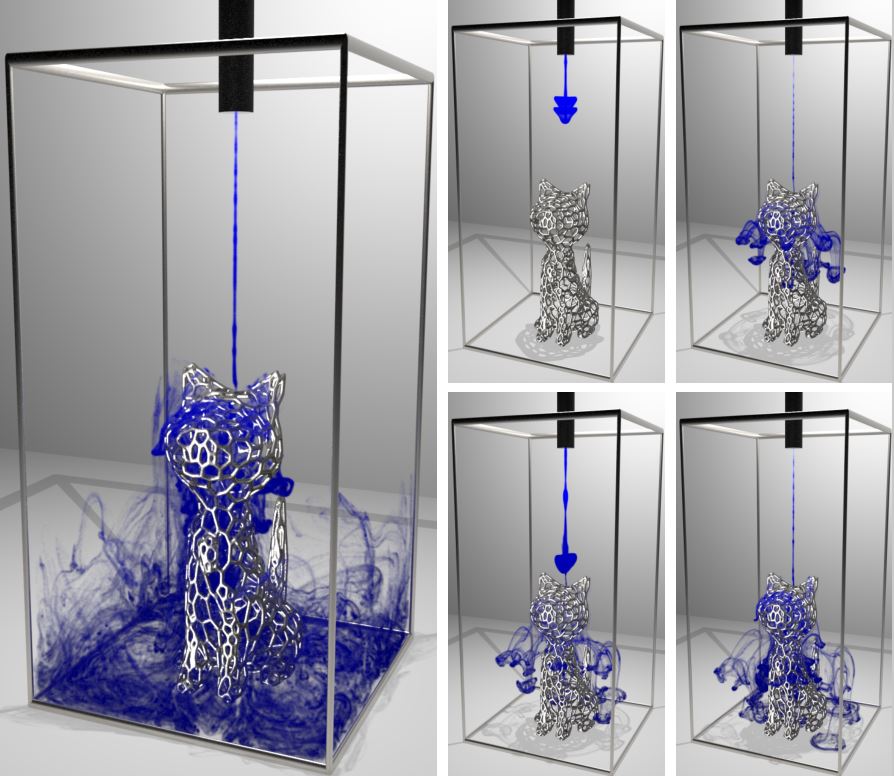}
\caption{Ink dropping onto an hollow cat: the ink interacts with the hollow cat and form complex patterns as the ink passes through the hollow cat, illustrating the robustness of our method.}
\label{fig:cat}
\end{minipage}
\end{figure*}

\section{Numerical Implementation}\label{sec:numerial_calculation}
\subsection{Fluid Calculation}
We employ the hybrid grid-particle method to perform numerical computations for the time-split scheme detailed in \autoref{sec:time_split}.  Particles are used to track flow maps and compute the mapped velocity, and the grid is used to perform gradient and other differential operations. We use subscripts \(q\) and \(i\) to distinguish between quantities on particles and quantities on the grid, respectively. 
 Each fluid particle carries particle initial velocity \(\mathbf{u}_{s,q}\), \revise{Jacobian} of backward flow map $\mathcal{T}_{s \to r}$, \revise{Jacobian} of forward flow map $\mathcal{F}_{s \to r}$ and the force path integral $\Gamma_{s\to r,q}$.  $\mathbf{x}_i$ and $\mathbf{x}_q$ denote positions of grid and particles respectively and in the processes of Particle-to-Grid Transfer and Grid-to-Particle Transfer, we use the quadratic kernel $w(\cdot)$ from \cite{jiang2016material} with denoting $w_{iq} = w(\mathbf{x}_q-\mathbf{x}_i)$.


\paragraph{\textbf{(1) Advection of Flow Maps}}  In \cite{deng2023fluid}, $\mathbf{x}_{r,q}$, $\mathcal{T}_{r \to s,q}$ and $\mathcal{F}_{s \to r,q}$ are advected using the Runge-Kutta 4th order method to solve $\frac{D\mathbf{x}_{r,q}}{Dt}=\mathbf{u}$ and \autoref{equ:advection_FT}. The velocity used to evolve $\mathbf{x}_{r,q}$, $\mathcal{T}_{r \to s,q}$ and $\mathcal{F}_{s \to r,q}$ can either be the velocity $\mathbf{u}_{s',i}$ from the previous step or the midpoint time velocity $\mathbf{u}_r^{\text{mid}}$ calculated according to Algorithm 2 in \cite{deng2023fluid}. Following \cite{deng2023fluid}, we use the midpoint time velocity to evolve $\mathbf{x}_{r,q}$, $\mathcal{T}_{r \to s,q}$ and $\mathcal{F}_{s \to r,q}$.

\paragraph{\textbf{(2) Calculation of $\frac{1}{2}(\nabla|\mathbf{u}_{s'}|^2)(\mathbf{x}_q(r))$}} 
The conversion from long-range mapped velocity to advected velocity occurs on the particles. However, the particles do not carry the quantity $\frac{1}{2}(\nabla|\mathbf{u}_{s'}|^2)(\mathbf{x}_q(r))$ in \autoref{equ:long_to_advect}. Therefore, it is necessary to compute the gradient $\revise{\frac{1}{2}(\nabla|\mathbf{u}_{s'}|^2)(\mathbf{x}_q(r))}$ at the particle positions through interpolation, using the final velocity $\mathbf{u}_{s'}$ from the previous step on the grid.
\begin{equation}\label{equ:p2g_grad_u}
    \frac{1}{2}(\nabla|\mathbf{u}_{s'}|^2)(\mathbf{x}_q(r))=\sum_{i}\frac{1}{2}|\mathbf{u}_{s',i}|^2\nabla w_{iq}.
\end{equation}

\paragraph{\textbf{(3) Particle-to-Grid Transfer}}  After calculating the mapped velocity on particles $\mathbf{u}_{s\to r,q}^M=\mathcal{T}_{r\to s,q}^T\mathbf{u}_{s,q}$,  the mapped velocity is converted to short-range advected velocity on particles based on \autoref{equ:long_to_advect}.
Then, the short-range advected velocity will be interpolated to the grid by APIC, similar to \cite{sancho2024impulse} as
\begin{equation}\label{equ:p2g_short_numerical}
    \mathbf{u}_{s'\to r,i}^A=\sum_{q}w_{iq} (\mathbf{u}_{s\to r,q}^A+A_q(\mathbf{x}_i-\mathbf{x}_q))/\revise{\left(\sum_{q}w_{iq}\right)},    
\end{equation}
where affine matrix $A_q$ is calculated during Grid-to-\revise{Particle} Transfer.

\paragraph{\textbf{(4) Volume Fraction and Sediment Velocity}} When solving the Poisson \autoref{equ:mixture_poisson} and computing the drag force, the velocity field of sediment and the volume fractions of sediment and fluid on the grid are required. The velocity field and volume fractions of sediment on the grid are given by: 
\begin{equation}\label{equ:laden_on_grid}
\begin{aligned}
    \epsilon^s_i=\frac{1}{\Delta x^3}\sum_{p} w_{ip}\frac{m_p}{\rho^s},\quad \revise{\mathbf{v}_i= \sum_{p}w_{ip}m_p\mathbf{v}_p/\left(\sum_{p}m_pw_{ip}\right)},
\end{aligned}
\end{equation}
where $\Delta x$ represents the size of \revise{a grid cell}. \revise{Here, the particle-to-grid process of velocity is consistent with the particle-to-grid process of velocity in MPM \cite{jiang2016material}. } Then volume fractions of fluid is calculated as $\epsilon^f_i=1-\epsilon^s_i$.

\paragraph{\textbf{(5) Force Calculation on Grid}} On the grid, we need to compute the viscous force and drag force.  The viscous force calculated by the finite difference \revise{using the six-point stencil} on the grid is given by:
\begin{equation}\label{equ:viscous_calculation}
    \revise{\left[\frac{\mu}{\rho^f}\Delta \mathbf{u}\right]_{i}}= \frac{\mu\sum_{j\in N_i}(\mathbf{u}^A_{s'\to r,j}-\mathbf{u}^A_{s'\to r,i})}{\rho^f|N_i|\Delta \mathbf{x}^2},
\end{equation}
where $N_i$ represents the neighboring \revise{cells} of $i$, in three-dimensional space, $|N_i|=6$. While the drag force is first computed for sediment particles $\mathbf{f}_{\text{drag},p}$ as \autoref{sec:sediment_calculation}, then utilizing \autoref{equ:drag_force_fluid}, it is computed on the grid through the following process:
\begin{equation}\label{equ:drag_force_fluid_numerical}
    \mathbf{f}^f_{\text{drag},i}= -\frac{1}{\rho^f \Delta x^3} \sum_{p} \mathbf{f}_{\text{drag},p} w_{ip}.
\end{equation}
  After calculating the viscous force and drag force, $\mathbf{u}^*$ is calculated by exerting these forces on $\mathbf{u}^A$:
  \begin{equation}
    \mathbf{u}_i^*=\mathbf{u}^A_{s'\to r,i}+\revise{\left(\left[\frac{\mu}{\rho^f}\Delta \mathbf{u}\right]_{i}+\frac{1}{\rho^f\epsilon^f}\mathbf{f}^f_{\text{drag},i}\right)}\Delta t. 
  \end{equation}
\paragraph{\textbf{(6) Pressure Projection}} Similar to \cite{gao2018animating}, on the grid, we solve the discretized variable coefficient Poisson equation \autoref{equ:mixture_poisson} with a multigrid preconditioner.  
After solving the Poisson equation, $\mathbf{u}_r$ is calculated as $\mathbf{u}_{r,i}=\mathbf{u}^A_{s'\to r, i}-\frac{1}{\rho^f}\nabla p_i.$

\paragraph{\textbf{(7) Grid-to-Particle Transfer}} During the Grid-to-Particle transfer process, $p_i$ and  $\gamma_i$ need to be interpolated to particles and then used to update $\Gamma_{s\to r,q}$ by \autoref{equ:update_Gamma} as:
\begin{equation}\label{equ:g2p_update_Gamma}
    \Gamma_{s\to r,q}=\Gamma_{s\to s',q}+\sum_i w_{iq}\gamma_i-\frac{1}{\rho^f}\sum_i p_i\nabla w_{iq} + \frac{1}{2}(\nabla |\mathbf{u}_{s'}|^2)(\mathbf{x}_q(r)).
\end{equation}
The affine matrix in APIC should be calculated for Particle-to-Grid process next step:
\begin{equation}\label{equ:affine_matrix_calculation}
    A_q=\sum_{i}\mathbf{u}_{r,i}\nabla w_{iq}.
\end{equation}

\begin{figure}
\centering
\begin{minipage}[t]{1.0\linewidth}
\centering
\includegraphics[width=1.0\linewidth]{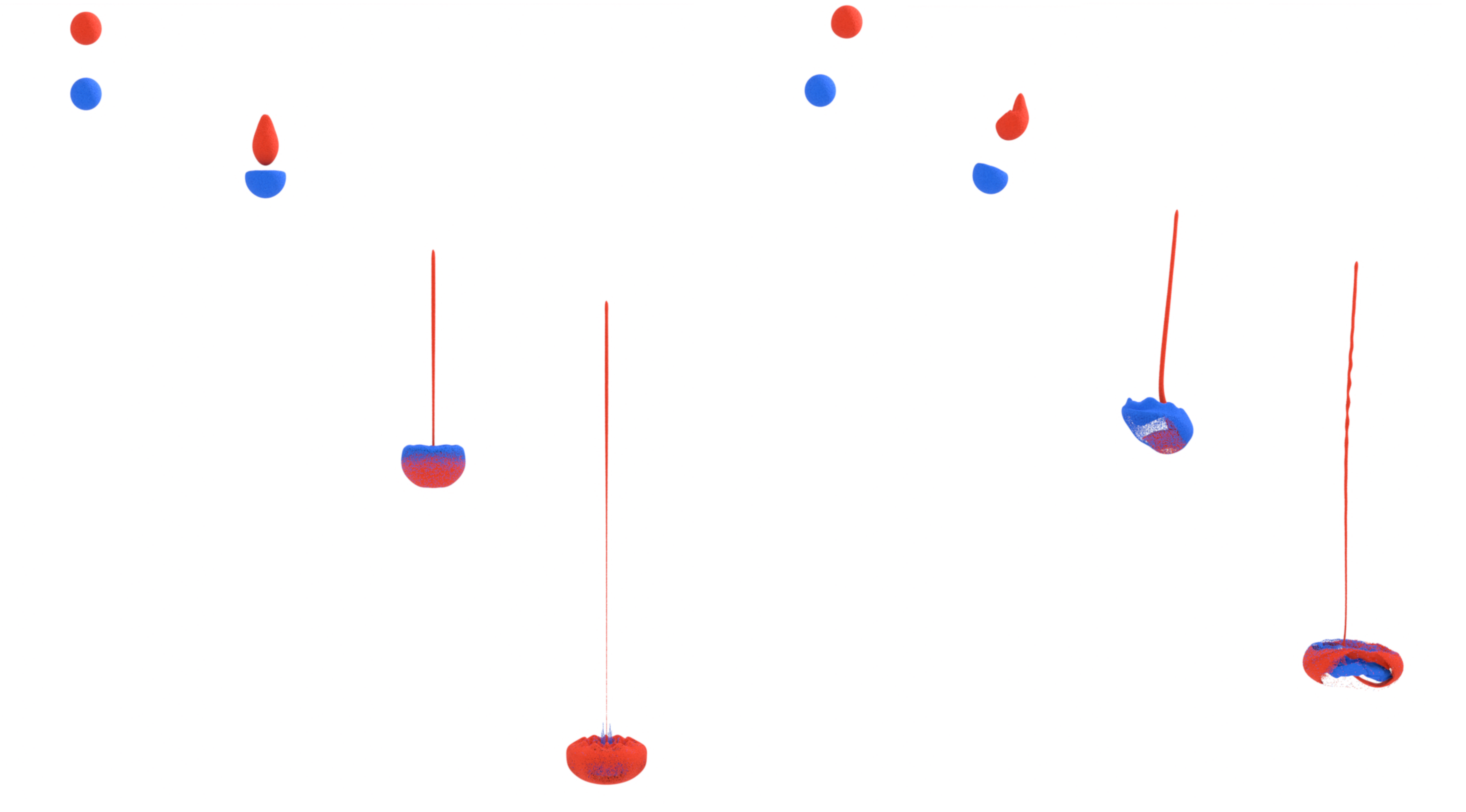}
\caption{Pass-Through of Two Drops. When two vertically aligned sediment drops fall simultaneously (left), the lower drop flattens into an oblate shape and the upper drop elongates into a prolate shape. The upper drop then catches up with and passes through the lower drop, and then two drops begin to mix. 
 Our result is consistent with Figure 14 in the real experiments from \cite{machu2001coalescence}.  When two \revise{horizontally offset drops} fall \revise{ (}right), the upper drop moves into vertical alignment with the lower drop, and then, similar to vertically aligned drops, catches up, penetrates, and mixes with the lower drop, which is consistent with Figure 16 in the real experiments from \cite{machu2001coalescence}. }
\label{fig:two_drop_low_re}
\end{minipage}
\begin{minipage}[t]{1.0\linewidth}
\centering
\includegraphics[width=1.0\textwidth]{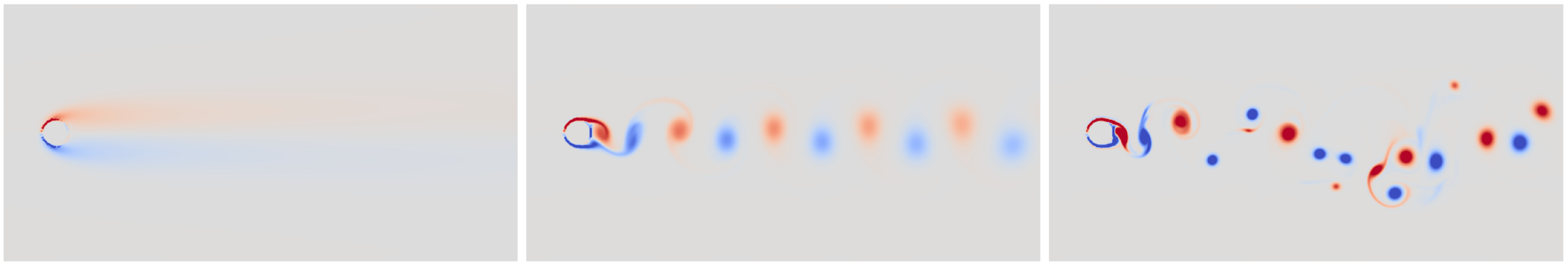}
 \caption{Kármán vortex shedding: Re=25, 250, and 2500.}
 \label{fig:karman}
\end{minipage}
\end{figure}
\subsection{Sediment Calculation}\label{sec:sediment_calculation}
\paragraph{Implicit Scheme} Following \cite{sagong2015simulating}, to improve stability, we use implicit Euler to compute \autoref{equ:sediment_dynamics} for updating particle velocities, where the numerical calculation form of \autoref{equ:sediment_dynamics} and \autoref{equ:drag_force_sediment} is written as:
\begin{equation}\label{equ:numerical_sediment_dynamics}
    \begin{aligned} 
    \frac{\mathbf{v}_p(r)-\mathbf{v}_p(s')}{\Delta t}&=\revise{\left(\frac{\rho^s-\rho^f}{\rho^s}\right)}g+\frac{6\pi\mu r_p}{m_p}(\mathbf{u}_r(\mathbf{x}_p)-\mathbf{v}_p(r)),\\
    \mathbf{f}_{\text{drag},p} &= \frac{6\pi\mu r_p}{m_p}(\mathbf{u}_r(\mathbf{x}_p)-\mathbf{v}_p(r)),
    \end{aligned}
\end{equation}
where \( \mathbf{u}_r(\mathbf{x}_p) \) is computed by interpolating the fluid velocity on the grid $\mathbf{u}_r(\mathbf{x}_p)=\sum_i w_{ip} \mathbf{u}_i$.  After updating the particle velocities, the particle positions are updated as: \( \mathbf{x}_p(r) = \mathbf{x}_p(s') + \mathbf{v}_p(s') \Delta t \).

\paragraph{Particle Cluster} When facing a physical scenario that includes a large number of sediment particles, computing the state of each particle using \autoref{equ:numerical_sediment_dynamics} is computationally too expensive. To address this issue, following \cite{bosse2005numerical}, we introduce particle clusters. Each particle cluster represents a group of \(N\) \revise{sediment} particles as a whole, assuming that the motion of these \(N\) particles is identical and described by the motion of the cluster. A particle cluster's motion follows \autoref{equ:numerical_sediment_dynamics}. When using clusters, the only modifications needed are to change \autoref{equ:drag_force_fluid_numerical} to:
\begin{equation}
\mathbf{f}^f_{\text{drag},i}= -\frac{N}{\Delta x^3} \sum_{p} \mathbf{f}_{\text{drag},p} w_{ip},
\end{equation}
and change sediment volume fraction calculation in \autoref{equ:laden_on_grid} to
\begin{equation}
        \epsilon^s_i=\frac{N}{\Delta x^3}\sum_{p} w_{ip}\frac{m_p}{\rho_s}.
\end{equation}
\paragraph{Boundary Treatment} We represent solid obstacles using a level set function \(L(\mathbf{x})\). After updating the position of any particle \(q\), we check \(L(\mathbf{x}_q)\). If \(L(\mathbf{x}_q) < 0\), we move \(\mathbf{x}_q\) along \(\nabla L(\mathbf{x}_q)\) until \(L(\mathbf{x}_q) \ge 0\), and update the particle velocity to \(\mathbf{v}_q - (\mathbf{v}_q \cdot \nabla L(\mathbf{x}_q)) \nabla L(\mathbf{x}_q)\).   

\section{Time Integration} 
We summarize our time integration scheme in Algorithm \autoref{alg:simulation_scheme}.  
%

\begin{algorithm}[h]
\caption{Laden Flow on Particle Flow Map}
\label{alg:simulation_scheme}
\begin{flushleft}
        \textbf{Initialize:} $\mathbf{u}_{i}, \mathbf{v}_p$ to initial velocity; $\mathcal{T}_{s \to s}$, $\mathcal{F}_{s \to s}$ to $\bm{I}$
\end{flushleft}
\begin{algorithmic}[1]
\For{$k$ in total steps}
\State $j \gets k \Mod {n^L}$;
\If{$j$ = 0}
\State Set initial time $s$ to now;
\State Uniformly distribute fluid particles;
\State Reinitialize $\mathcal{T}_{s\to s}$,$\mathcal{F}_{s\to s}$ to $\bm{I}$;
\State Reinitialize velocity $\mathbf{u}_{s,q}$ by interpolating from $\mathbf{u}_{s,i}$;
\EndIf
\State Compute $\Delta t$ with $\mathbf{u}_i$, $\mathbf{v}_p$ and the CFL number;

\State Estimate midpoint velocity $\bm{u}_i^\text{mid}$; 

\State Advect $\bm x_q$, $\mathcal{T}_{r\to s}$, $\mathcal{F}_{s\to r}$ with $\bm{u}_i^\text{mid}$ and $\Delta t$; 

\State Calculate mapped velocity $\mathbf{u}_{s\to r,q}^M$ and convert to short-range advected velocity $\mathbf{u}_{s'\to r,q}^A$; \hfill $\triangleright$ eq. \ref{equ:long_to_advect}

\State Compute $\frac{1}{2}(\nabla |\mathbf{u}_{s'}|^2)(\mathbf{x}_q(r))$ by interpolation;  \hfill $\triangleright$ eq. \ref{equ:p2g_grad_u}

\State Compute $\mathbf{u}_{s'\to r,i}^A$ by transfer $\mathbf{u}_{s'\to r,q}^A$ to grid; \hfill $\triangleright$ eq. \ref{equ:p2g_short_numerical}

\State Compute volume fraction $\epsilon_i^s$,$\epsilon_i^f$ and sediment velocity $\mathbf{u}_i^s$ on the grid;  \hfill $\triangleright$ eq. \ref{equ:laden_on_grid}

\State Compute viscous force;  \hfill $\triangleright$ eq. \ref{equ:viscous_calculation}.

\State Compute drag force and update sediment state; \hfill $\triangleright$ eq. \ref{equ:numerical_sediment_dynamics}, \ref{equ:drag_force_fluid_numerical}.

\State Compute $\mathbf{u}_i^*$;  \hfill $\triangleright$ eq. \ref{equ:compute_u_star}.

\State Compute pressure $p_i$ by solving Poisson equation and get final velocity $\mathbf{u}_{r,i}$ by projection; \hfill $\triangleright$ eq. \ref{equ:mixture_poisson}.

\State Update path integration $\Gamma_{s\to r,q}$ in Grid-to Particle Transfer and calculate the affine matrix $A_q$; \hfill $\triangleright$ eq. \ref{equ:g2p_update_Gamma},\ref{equ:affine_matrix_calculation}.

\EndFor{}
\end{algorithmic}
\end{algorithm}

\section{Results and Discussion}
\paragraph{Validation} 
We first compare our simulation results with real-world phenomena to verify the accuracy of our method.
\textbf{(1) Kármán Vortex Shedding.} We validate the correctness of adding viscosity as described in \autoref{sec:effect_of_force} through 2D Kármán vortex shedding experiments. As illustrated in Fig.~\ref{fig:karman}, the phenomena of Kármán vortex shedding are obtained through our simulations at different Reynolds numbers $Re=25, 250,$ and $2500$.  \revise{At three different Reynolds numbers, our simulation results exhibited three distinct patterns: a laminar wake without a vortex street at low Re (25), a periodic vortex street at moderate Re (250), and turbulent mixing at high Re (2500), which is consistent with the physical experiments described in \cite{blevins1977flow}.}

\textbf{(2) Pass-through for Vertically Aligned Drop.} Fig.~\ref{fig:two_drop_low_re} shows the phenomenon called pass-through when two vertically aligned sediment drops fall under low   Reynolds number $Re=3$, which is consistent with Figure 14 in the real experiments from \cite{machu2001coalescence}. 
\textbf{(3) Pass-through for Horizontally Offset Drop.} Fig.~\ref{fig:two_drop_low_re} shows the pass-through phenomenon when two horizontally offset drops fall, which is consistent with Figure 16 in the real experiments from \cite{machu2001coalescence}.
\textbf{(4) Ink Torus Breakup Under Different Reynolds Number.}  When a sediment drop falls, it forms an ink torus, which disintegrates into several blobs consequently. According to \cite{bosse2005numerical}, the number of blobs from the torus increases with the Reynolds Number.  We use $Re=16.5, 18, 19.5, 21.0, 30.0$ and observe that the number of blobs gradually increases from 4 to 8, as shown in Fig.~\ref{fig:multi_re}.

{
\begin{figure}
\centering
\begin{minipage}[t]{1.0\linewidth}
\begin{minipage}[t]{1\linewidth}
\centering
\includegraphics[width=1.0\linewidth]{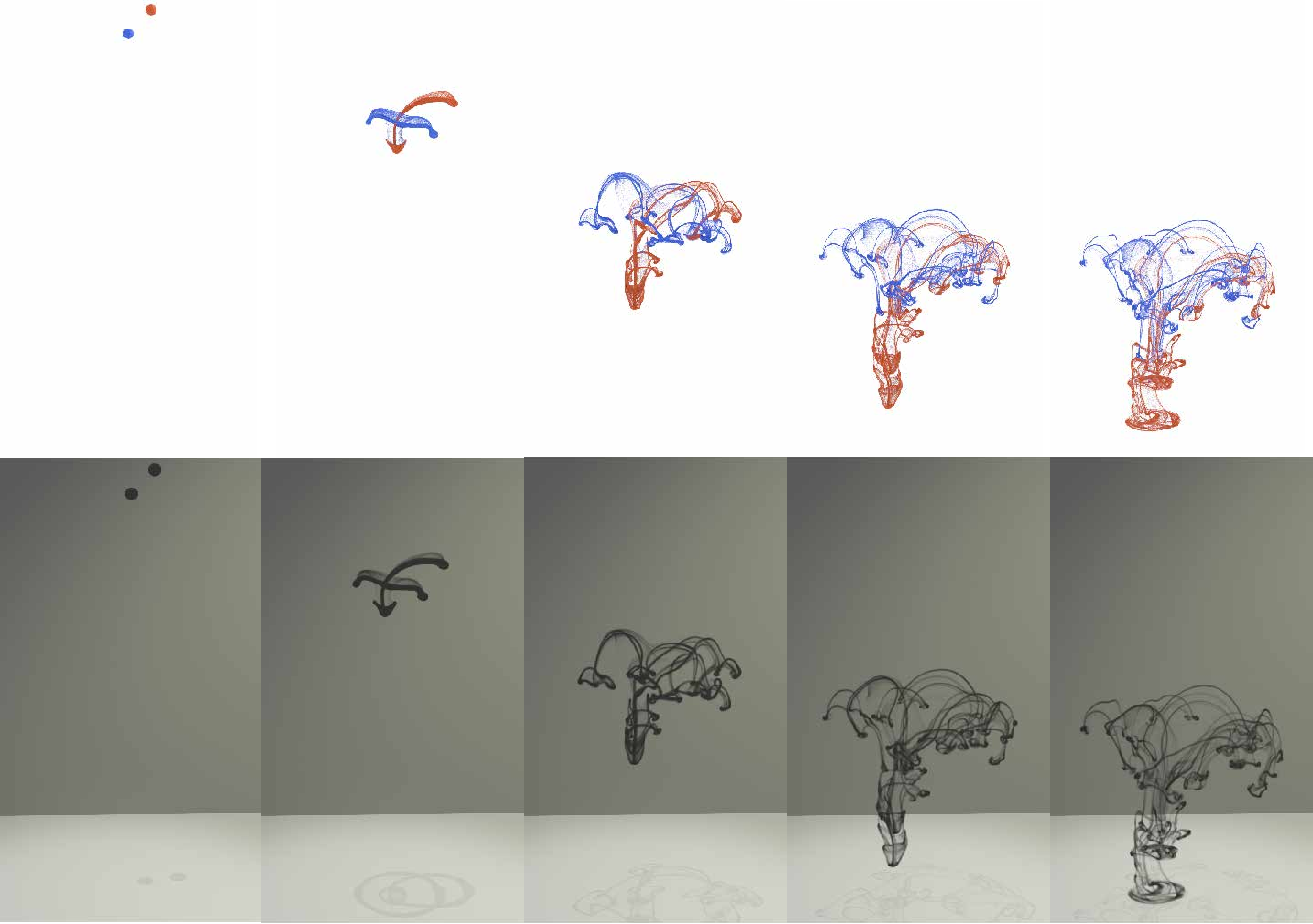}
\caption{Two Drop Interaction. At a high Reynolds Number, two drops interact and form a complex pattern.}
\label{fig:two_drop_high_re_non_vertical}
\end{minipage}
\begin{minipage}[t]{1\linewidth}
\includegraphics[width=1.0\linewidth]{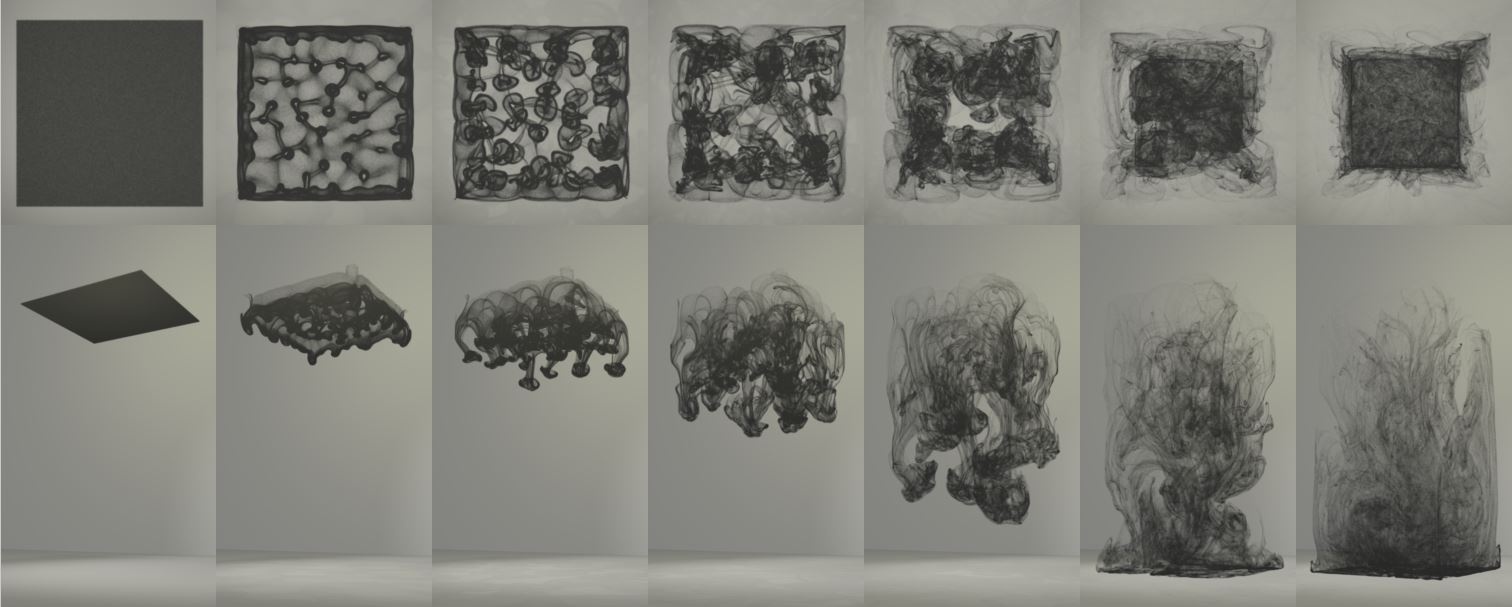}
\caption{RT Instability. The initially thin layer of ink at the top of the water tank gradually becomes unstable. Through the process of coalescence, it forms many small ink drops. These aggregated ink drops begin to fall and take on the classic figure shape of RT instability.  }
\label{fig:one_layer}
\end{minipage}
\end{minipage}
\end{figure}
}

\paragraph{Comparison.}

\revise{Next, we demonstrate the effectiveness of our method by performing three experiments against four laden flow solvers, which use Semi-Lagrangian \cite{stam1999stable}, APIC \cite{jiang2015affine}, Bimocq\cite{qu2019efficient}, and the basic flow-map method \cite{sato2017long} to compute the fluid portion, respectively.  The laden flow solver with the Semi-Lagrangian method of solving fluid on grids is based on \cite{sagong2015simulating}.} 
\textbf{(1) Ink Torus Breakup Comparison.} Fig.~\ref{fig:two_drop_leapfrog} compares the disintegration of the merged torus formed by the descent of two ink droplets, computed using different methods at \revise{$Re=15$}. 
\textbf{(2) Two Ink Drops Oblique Collision.} Fig.~\ref{fig:three_drop_collision}(above) compares the reconnection of ink torus when two ink tori collide with three methods. 
\textbf{(3) Three Ink Drops Oblique Collision.} Fig.~\ref{fig:three_drop_collision} (below) compares the reconnection of ink torus when three ink tori collide with three methods.

\begin{table*}[t]
\caption{\revise{The catalog of all our simulation examples. \#Fluid is the number of fluid particles per cell and \#Sediment is the total number of sediment particle clusters}}
\centering\small
\begin{tabularx}{\textwidth}{p{5.5cm} | p{2cm} | p{2cm} | Y | Y | Y |Y}
\hlineB{3}
\revise{Name} & \revise{Figure} & \revise{Resolution}  & \revise{\#Fluid} & \revise{\#Sediment} & \revise{Time (sec /substep)} &\revise{Memory Cost (GB)} \\
\hlineB{2.5}
\revise{Kármán Vortex Shedding} & \revise{Figure~\ref{fig:karman}} & \revise{512 $\times$ 256}   & \revise{16} & \revise{N/A} & \revise{0.91} &\revise{0.43}  \\
\hlineB{2}
\revise{Ink Torus Breakup} & \revise{Figure~\ref{fig:single_drop}} & \revise{128 $\times$ 256 $\times$ 128}   & \revise{8} & \revise{$1.0\times 10^6$} & \revise{1.84} &\revise{11.49}\\
\hlineB{2}
\revise{Ink Torus Breakup(Under Different Re)} & \revise{Figure~\ref{fig:multi_re}}& \revise{128 $\times$ 256 $\times$ 128}   & \revise{8} & \revise{$1.0\times 10^6$} & \revise{1.84} &\revise{11.49}\\
\hlineB{2}
\revise{Pass-Through of Two Drops (Vertically Aligned)} & \revise{Figure~\ref{fig:two_drop_low_re}(left)} & \revise{128 $\times$ 256 $\times$ 128}   & \revise{8} & \revise{$1.0\times 10^6$} & \revise{1.83} &\revise{11.49}\\
\hlineB{2}
\revise{Pass-Through of Two Drops (Horizontally Offset)} & \revise{Figure~\ref{fig:two_drop_low_re}(right)} & \revise{128 $\times$ 256 $\times$ 128}   & \revise{8} & \revise{$1.0\times 10^6$} & \revise{1.81} &\revise{11.49}\\
\hlineB{2}
\revise{Two Ink Drops Oblique Collision Comparison} & \revise{Figure~\ref{fig:three_drop_collision}(above)} & \revise{128 $\times$ 256 $\times$ 128}   & \revise{8} & \revise{$1.0\times 10^6$} & \revise{1.85} &\revise{11.49}\\
\hlineB{2}
\revise{Three Ink Drops Oblique Collision Comparison} & \revise{Figure~\ref{fig:three_drop_collision}(below)} & \revise{128 $\times$ 256 $\times$ 128}   & \revise{8} & \revise{$1.5\times 10^6$} & \revise{1.83} &\revise{11.52}\\
\hlineB{2}
\revise{Ink Torus Breakup Comparison} & \revise{Figure~\ref{fig:two_drop_leapfrog}} & \revise{128 $\times$ 256 $\times$ 128}   & \revise{8} & \revise{$1.0\times 10^6$} & \revise{1.84} &\revise{11.49}\\
\hlineB{2}
\revise{Two Drop Interaction} & \revise{Figure~\ref{fig:two_drop_high_re_non_vertical}} & \revise{128 $\times$ 256 $\times$ 128}   & \revise{8} & \revise{$1.0\times 10^6$} &\revise{1.81} &\revise{11.49}\\
\hlineB{2}
\revise{Dripping} & \revise{Figure~\ref{fig:inject1}} & \revise{128 $\times$ 256 $\times$ 128}   & \revise{8} & \revise{$2.0\times 10^6$} & \revise{1.85} &\revise{11.55}\\
\hlineB{2}
\revise{RT Instability} & \revise{Figure~\ref{fig:one_layer}} & \revise{128 $\times$ 256 $\times$ 128}   & \revise{8} & \revise{$4.0\times 10^6$} & \revise{1.94} &\revise{11.66}\\
\hlineB{2}
\revise{One Ink Drop Passing Sphere Obstacle} & \revise{Figure~\ref{fig:small_sphere}} & \revise{128 $\times$ 256 $\times$ 128}   & \revise{8} & \revise{$1.0\times 10^6$} & \revise{1.99} & \revise{11.49}\\
\hlineB{2}
\revise{Chopstick Stirring Ink} & \revise{Figure~\ref{fig:stir} }& \revise{128 $\times$ 256 $\times$ 128}   & \revise{8} & \revise{$1.0\times 10^6$} & \revise{2.06} &\revise{11.49}\\
\hlineB{2}
\revise{Nine Ink Drops Passing Porous Obstacle} & \revise{Figure~\ref{fig:porus}} & \revise{128 $\times$ 256 $\times$ 128}   & \revise{8} & \revise{$3.0\times 10^6$} & \revise{2.32} &\revise{11.49}\\
\hlineB{2}
\revise{Ink Dropping onto an Armadillo} & \revise{Figure~\ref{fig:armadillo}} & \revise{128 $\times$ 256 $\times$ 128}   & \revise{8} & \revise{$4.0\times 10^6$} & \revise{2.21} & \revise{11.66}\\
\hlineB{2}
\revise{ Ink Dropping onto a Hollow Cat} & \revise{Figure~\ref{fig:cat}} & \revise{128 $\times$ 256 $\times$ 128}   & \revise{8} & \revise{$2.0\times 10^6$} & \revise{2.06} &\revise{11.55} \\
\hlineB{3}
\end{tabularx}
\vspace{5pt}
\label{tab: examples_table}

\end{table*}

\paragraph{Examples} 
In our experiment, the fluid part is calculated using a $\revise{128\times128\times256}$ grid, with each grid cell containing 8 fluid particles. For the sediment part, we use 1,000,000 to 4,000,000 particle clusters. $CFL=0.5$ is used for calculate the $\Delta t$. We use Taichi \cite{hu2019taichi} for our implementation, and experiments are run on Tesla A100 GPUs and each substep takes approximately 2.0 seconds. \revise{The details of memory cost and runtime are presented in Table 1.}
\textbf{(1) Two Drop Interaction.} In Fig.~\ref{fig:two_drop_high_re_non_vertical}, at $Re=30$, two horizontally offset drops interact as they fall, forming a complex pattern composed of ink rings and filaments. 
\textbf{(2) \revise{Rayleigh-Taylor (RT)} Instability.} Fig.~\ref{fig:one_layer} shows a thin layer of ink turns unstable by gravity and forms small ink drops, leading to the classic figure shape of RT instability. 
\textbf{(3) Chopstick Stirring Ink.} As shown in Fig.~\ref{fig:stir}, under the action of stirring, the ink initially dropped into the water tank gradually becomes thin and forms many layers, pulling out many fine threads. As the stirring time increases, the ink gradually spreads throughout the entire water tank. 
\textbf{(4) One Ink Drop Passing Sphere Obstacle.} In Fig.\ref{fig:small_sphere}, as an ink drop descends past a spherical obstacle, it flows over the surface of the sphere, and part of the ink coalesces into small drops near the bottom of the ink and then separating, while the remaining ink continues to flow along the surface of the sphere and gathers at the bottom. Under the influence of gravity, these gathered ink drops then fall downward. During falling, the aggregated ink continuously forms torus shapes and disintegrates into smaller blobs. 
\textbf{(5) Nine Ink Drops Passing Porous Obstacle.} As shown in Fig.\ref{fig:porus}, when 9 drops of ink fall through a porous obstacle composed of several cylindrical structures, the ink drips down from the gaps between cylinders, forming many falling drops. These falling drops interact, creating a complex pattern. 
\textbf{(6) Dripping.} At the top of the water tank, we intermittently drip ink into the tank. As shown in Fig.~\ref{fig:inject1} the ink gradually falls and impacts at the bottom of the tank. During the ink falling, a long tail forms between it and the dropper. This tail will bulge and break up, forming many small suspension drops, which is consistent with the real photographs in Figure 6 of \cite{machu2001coalescence}. As the small suspension drops continue to fall, they develop into tori, which is consistent with Figure 9. in \cite{rogers1858art}. 
\textbf{(7) Ink Dripping onto an Obstacle.} In Fig~\ref{fig:armadillo} and Fig.~\ref{fig:cat}, we dripped ink onto complex solids to demonstrate the phenomena of ink and complex solid interaction, proving our method's robustness.

{
\begin{figure*}
\centering
\hspace{1cm}
\begin{minipage}[t]{1\linewidth}
\includegraphics[width=1\linewidth]{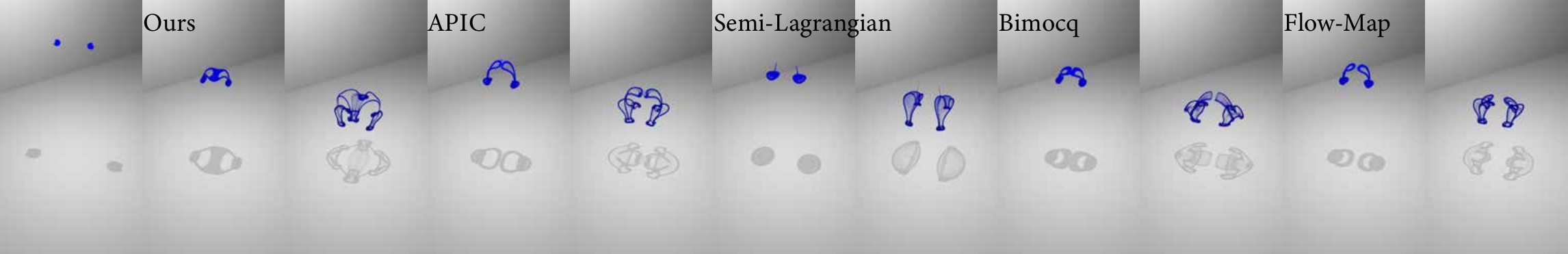}

\includegraphics[width=1\linewidth]{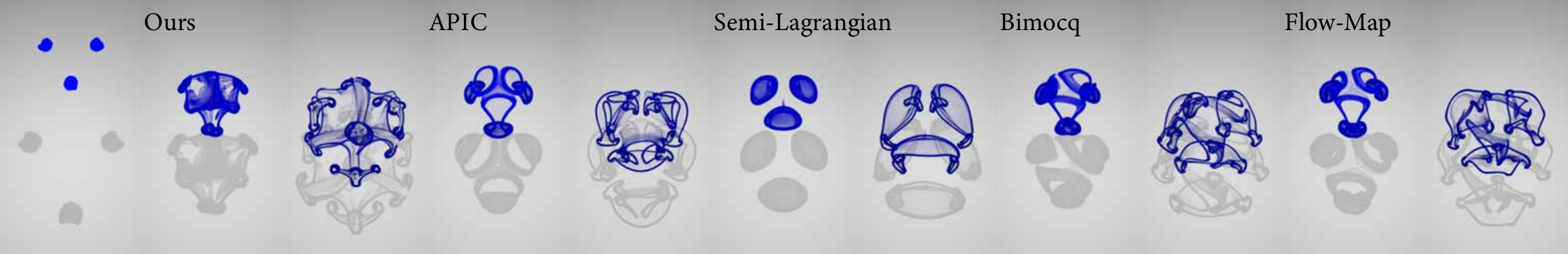}
\caption{
Ink Drops Oblique Collision Comparison. When two Ink Drops collide obliquely (above), the tori formed by the two drops exhibits ring reconnection similar to vortex ring collisions. \revise{Compared to APIC, Semi-Lagrangian, Bimocq and basic flow-map methods}, our method better maintains the vortex results, achieving ink ring reconnection uniquely.  When three Ink Drops collide obliquely (below), \revise{only our method achieves ink ring reconnection compared to APIC, Semi-Lagrangian, Bimocq and basic flow-map methods}, forming a new small ink ring at the center.}
\label{fig:three_drop_collision}
\end{minipage}
\begin{minipage}[t]{1\linewidth}
\includegraphics[width=1\linewidth]{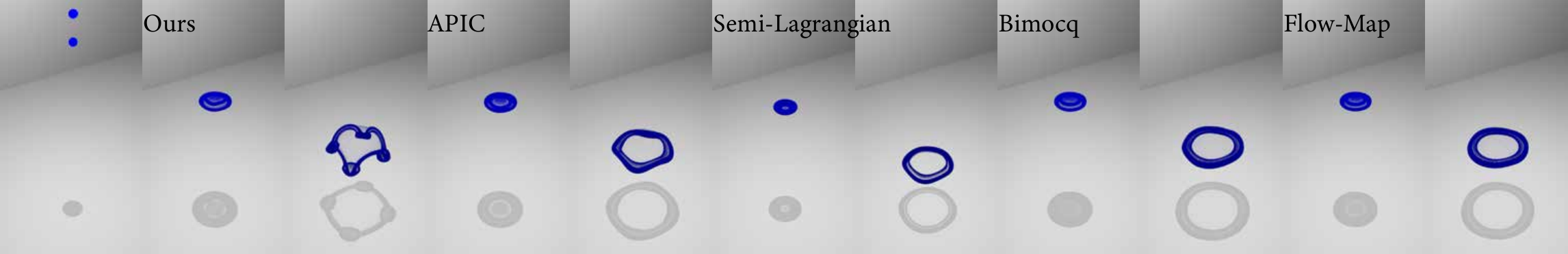}
\caption{ Ink
Torus Breakup Comparison. At \revise{$Re=15$}, when two vertically aligned drops fall, after they undergo the process of pass-through, mixing and forming a torus, the torus forming by two drops disintegrates into more rings as they continue to fall. Due to our method's lower numerical dissipation, the formed torus disintegrates into more rings more effectively.}
\label{fig:two_drop_leapfrog}
\end{minipage}

\end{figure*}
}

\revise{
\paragraph{Discussion}
Our particle-laden flow map model extends the simulation scope of traditional impulse/covector methods (e.g., \cite{nabizadeh2022covector,deng2023fluid,feng2022impulse} to the realm of viscosity-driven and particle-laden fluid phenomena. This scope extension is due to our method's new capability of tackling forces such as fluid viscosity and laden particle drag, which play an important role in producing physically-based vortical flow structures. Traditional impulse/covector methods such as \cite{nabizadeh2022covector,deng2023fluid} solve the Euler equation for inviscid incompressible flow that does not consider viscosity. Moreover, the force that sediment particles exert on the fluid cannot be incorporated into the traditional impulse/covector fluid models, limiting their modeling of interactions between sediment particles and their surrounding fluid. Last, the fluid portion of particle-laden fluid flow requires solving a variable-coefficient Poisson equation, which makes it difficult for solvers relying on solving a constant-coefficient Poisson system (e.g., the advection-reflection method \cite{zehnder2018advection}) to apply. 
}

\section{Limitations and Future Work}
In summary, we have enhanced the flow map method to accommodate forces beyond pressure and introduced a novel framework for simulating ink as a particle-laden flow using particle flow maps. The most significant limitation of our framework is that it only considers the interaction forces between the fluid and sediment, neglecting interactions between sediment particles. This limitation prevents the handling of phenomena such as sediment accumulation. We plan to incorporate interactions between sediment particles into our framework to enable the simulation of a \revise{broader} range of laden flow phenomena. We also consider implementing adaptive particles to achieve larger-scale laden flow effects such as sand storms. 

\section*{Acknowledgements}
We express our gratitude to the anonymous reviewers for their insightful feedback. Georgia Tech authors also acknowledge NSF IIS \#2433322, ECCS \#2318814, CAREER \#2433307, IIS \#2106733, OISE \#2433313, and CNS \#1919647 for funding support. We credit the Houdini education license for video animations.

\appendix
\section{Derivation of Eq.10}\label{sec:Covector_Derivation}
Similar to \cite{nabizadeh2022covector}, we reformulated \autoref{equ:fluid_momentum_gamma} as $\frac{\partial \mathbf{u}}{\partial t}+(\mathbf{u}\cdot \nabla)\mathbf{u}+\nabla {\mathbf{u}}^T\cdot \mathbf{u}=\gamma -\frac{1}{\rho^f}\nabla \revise{\left(p 
 +\frac{1}{2}\rho^f|{\mathbf{u}}|^2\right)}$ and expressed it in covector form using the Lie derivative 
 \begin{equation}
     \revise{\left(\frac{\partial}{\partial t}+L_{\mathbf{u}}\right){\mathbf{u}}^\flat=\gamma^\flat-d\lambda}
 \end{equation}
 By integrating this equation of Lie derivative form, we obtain
 \begin{equation}
 \begin{aligned}
    \mathbf{u}^\flat_r&=\revise{\Psi_{r \to s}^*\mathbf{u}^\flat_s+\int_s^t (\Phi_{s \to \tau}\circ \Psi_{r \to s} )^*(\gamma^\flat_\tau-d\lambda_\tau)d\tau}\\
    & = \revise{\Psi_{r \to s}^*\mathbf{u}^\flat_s+\Phi_{s \to \tau}^*\int_s^t \Psi_{r \to s}^*(\gamma^\flat_\tau-d\lambda_\tau)d\tau}
 \end{aligned}
 \end{equation}
where $\Psi_{r \to s}^*$ and $\Phi_{s \to \tau}^*$ are the pullbacks of the \revise{covector} induced by $\Psi_{r \to s}$ and $\Phi_{s \to \tau}$, respectively. Convert the above expression back to vector form, and noting that $\revise{\Psi_{r\to s}^*\mathbf{v}^\flat}$ and $\revise{\Phi_{s\to \tau}^*\mathbf{v}^\flat}$ corresponds to $\nabla \Psi_{r\to s}^T \mathbf{v}$ and $\nabla \Phi_{s\to \tau}^T \mathbf{v}$ respectively for arbitrary vector field $\mathbf{v}$.

\bibliographystyle{ACM-Reference-Format}
\bibliography{refs_ML_sim.bib, refs_INR.bib, refs_flow_map.bib, refs_simulation.bib, refs_laden.bib,bo.bib}


\pagebreak

\end{document}